\documentclass[a4paper,USenglish]{lipics}

\usepackage{cite}

\usepackage{array}
\usepackage{url}
\usepackage{amsthm}
\usepackage{amssymb}
\usepackage{wasysym,stmaryrd}
\usepackage[T1]{fontenc}
\usepackage{bussproofs}
\usepackage{tikz-qtree}
\usepackage{tikz}
\usepackage{graphicx}
\usepackage{cjhebrew}
\usepackage{cmll}

\newcommand{\msp}{\:\!}                        

\newcommand{\data}{\texttt{data}}
\newcommand{\ift}{\texttt{if}}
\newcommand{\Nil}{\texttt{Nil}}

\newcommand{\Koba}{\mathit{Kob}(\G,\APT)}
\newcommand{\KOname}{\mathit{KO}}
\newcommand{\KO}{\KOname(\mathcal{A})}
\newcommand{\KOnew}{$\begin{cjhebrew}.s\end{cjhebrew}$(\mathcal{A})}
\newcommand{\KOY}{\KOname_{fix}(\mathcal{G},\mathcal{A})}
\newcommand{\KOYnew}{$\begin{cjhebrew}.s\end{cjhebrew}$_{fix}(\mathcal{G},\mathcal{A})}
\newcommand{\tsadi}{\begin{cjhebrew}.s\end{cjhebrew}}

\newcommand{\G}{\mathcal{G}}

\newcommand{\rG}{\rightarrow_{\mathcal{G}}}
\newcommand{\valG}{[\![ \mathcal{G} ] \! ]}
\newcommand{\termG}{t(\mathcal{G})}

\newcommand{\KObox}{\boxbox}

\newcommand{\setkinds}{\mathcal{K}}
\newcommand{\kind}{\operatorname{kind}}

\newcommand{\arity}{\operatorname{ar}}
\newcommand{\order}{\operatorname{order}}

\newcommand{\interp}[1]{ [\![ #1  ]\!]}
\newcommand{\interpinf}[1]{ [\![ #1  ]\!]_{\superbang}}
\newcommand{\interpcol}[1]{ [\![ #1  ]\!]_{\superbang\hspace{-.5em}\superbang\hspace{-.5em}\superbang\hspace{-.5em}\superbang}}
\newcommand{\interpcoltest}[1]{ [\![ #1  ]\!]_{\superbang\hspace{-.47em}\superbang\hspace{-.47em}\superbang\hspace{-.5em}\superbang}}
\newcommand{\flechedimplication}{\rightarrow}
\newcommand{\KOYtranslation}[1]{\pi(#1)}

\newcommand{\APT}{\mathcal{A}}

\newcommand{\Adamic}{\textit{\textbf{Adamic}}(\mathcal{G},\mathcal{A})}
\newcommand{\Edenic}{\textit{\textbf{Edenic}}(\mathcal{G},\mathcal{A})}
\newcommand{\Edenicnew}{\textit{\textbf{Edenic}}^{new}(\mathcal{G},\mathcal{A})}

\newcommand{\Edenictranslation}[1]{\sigma(#1)}

\newcommand{\semantics}[1]{ || #1  ||}

\newcommand{\fixrule}{fix}

\newcommand{\superbang}{\lightning} 

\newcommand{\colorbang}{\superbang\hspace{-.5765em}\superbang\hspace{-.5765em}\superbang}

\newcommand{\Relinfinitary}{\underline{Rel}}
\newcommand{\tensor}{\otimes}

\newcommand{\morph}[1]{\stackrel{#1}{\longrightarrow}}

\newtheorem{proposition}{Proposition}

\title{Relational semantics of linear logic and higher-order model-checking}

\author[1]{Charles Grellois}
\author[2]{Paul-André Melliès}
\affil[1]{Laboratoires PPS \& LIAFA, Université Paris Diderot\\
  \texttt{grellois@pps.univ-paris-diderot.fr}}
\affil[2]{Laboratoire PPS, CNRS \& Université Paris Diderot\\
  \texttt{mellies@pps.univ-paris-diderot.fr}}
\authorrunning{C. Grellois and P.-A. Melliès}

\Copyright{Charles Grellois and Paul-André Melliès}

\subjclass{Dummy classification -- please refer to \url{http://www.acm.org/about/class/ccs98-html}}

\serieslogo{}
\volumeinfo
  {Billy Editor and Bill Editors}
  {2}
  {Conference title on which this volume is based on}
  {1}
  {1}
  {1}
\EventShortName{}
\DOI{10.4230/LIPIcs.xxx.yyy.p}

\begin{document}

\maketitle

\begin{abstract}
In this article, we develop a new and somewhat unexpected connection between higher-order model-checking and linear logic.
Our starting point is the observation that once embedded in the relational semantics of linear logic, the Church encoding
of a higher-order recursion scheme (HORS) comes together with a dual Church encoding of an alternating tree automata (ATA)
of the same signature. Moreover, the interaction between the relational interpretations of the HORS and of the ATA
identifies the set of accepting states of the tree automaton against the infinite tree generated by the recursion scheme.
We show how to extend this result to alternating parity automata (APT) by introducing a parametric version of the
exponential modality of linear logic, capturing the formal properties of colors (or priorities) in higher-order model-checking.
We show in particular how to reunderstand in this way the type-theoretic approach to higher-order model-checking developed by Kobayashi and Ong. We briefly explain in the end of the paper how this analysis driven by linear logic results in a new and purely semantic proof of decidability of the formulas of the monadic second-order logic for higher-order recursion schemes. 
\end{abstract}


\section{Introduction}
\label{section/introduction}
Thanks to the seminal works by Girard and Reynolds
on polymorphism and parametricity in the 1970s,
it has been recognized that every finite tree~$t$ on a given signature~$\Sigma$
can be seen alternatively as a simply-typed $\lambda$-term of an appropriate type depending on~$\Sigma$.
This correspondence between trees and $\lambda$-terms
is even bijective if one considers $\lambda$-terms
up to $\beta\eta$-equivalence, see for instance Girard~\cite{proofs-and-types}.
Typically, a finite tree~$t$ on the signature
\begin{equation}
\label{equation/signature}
\Sigma \quad = \quad \{\,a:2 \, , \, b:1 \, , \,  c:0 \, \}
\end{equation}
is the same thing under this Church encoding as a simply-typed $\lambda$-term~$t$ of type
\begin{equation}\label{equation/types-of-sigma}
(o\rightarrow o\rightarrow o)\rightarrow (o\rightarrow o) \rightarrow o\rightarrow o
\end{equation}
modulo $\beta\eta$-equivalence.
The idea underlying the correspondence is that every constructor~$a,b,c$
of the signature~$\Sigma$ should be treated as a variable
\begin{equation}\label{equation/types-of-a-b-c}
a \hspace{.5em} : \hspace{.5em} o \, \to \, o \, \to \, o
\quad\quad
b \hspace{.5em} : \hspace{.5em} o \, \to \, o
\quad\quad
c \hspace{.5em} : \hspace{.5em} o
\quad\quad
\end{equation}
where the number of inputs~$o$ in the type ${o\to \cdots\to o\to o}$ 
of the variable $a,b,c$ indicates the arity of the combinator.
An equally well-known fact is that this translation extends to infinite trees
generated by higher-order recursion schemes on the signature~$\Sigma$
if one extends the simply-typed $\lambda$-calculus with a fixpoint operator~~$Y$.
%
For example, the higher-order recursion scheme $\mathcal{G}$ on the signature~$\Sigma$ 
\begin{equation}\label{equation/hors}
\begin{array}{c}
\mathcal{G} \,\, = \,\,
\left\{\begin{array}{lcl}
 S &  \mapsto  & F\ a\ b\ c\\
 F\ x\ y \ z &  \mapsto  & x  \ (y \ z )\ (F\ x \ y \ (y\ z))
\end{array}\right.
\end{array}
\end{equation}
constructs the infinite tree
\begin{equation}\label{equation/value-tree}
\begin{tabular}{ccc}
$\valG$ & \quad  $=$ \quad \quad\quad
\raisebox{-6em}{\includegraphics[height=12.5em]{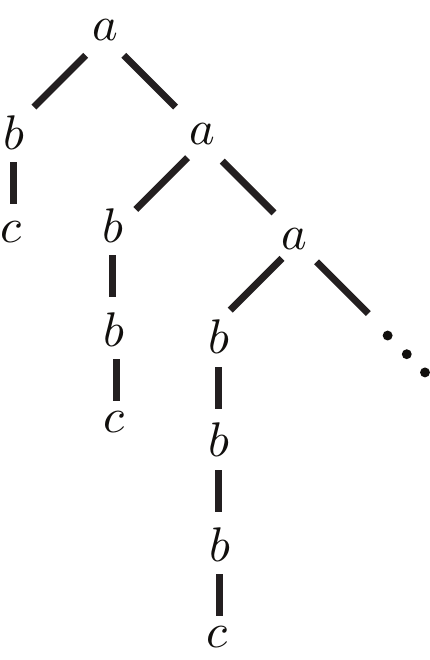}}
\end{tabular}
\end{equation}
and can be formulated as a $\lambda$-term of the same type~(\ref{equation/types-of-a-b-c})
as previously but defined in the simply-typed $\lambda$-calculus extended
with the fixpoint operator~$Y$:
\begin{equation}\label{equation/lambdaY-term}
\lambda a b c.\,\left(\left(Y\,\left(\lambda F.\left( \lambda xyz.\, x  \, (y \, z )\ (F\, x \, y \, (y\, z)) \right) \right) \right)\, a \, b\, c \right)
\end{equation}
%
A natural temptation is to study the correspondence between higher-order recursion schemes (\ref{equation/hors})
and simply-typed $\lambda$-terms with fixpoints (\ref{equation/lambdaY-term}) from the resource-aware point of view of linear logic.
%
%
%
Recall that the intuitionistic type~(\ref{equation/types-of-a-b-c})
is traditionally translated in linear logic as the formula
$$
A \quad = \quad !\, (\, \, !\, o \, \multimap \, {!\, o} \, \multimap o \, \, )\multimap \, {!}\, (\, \, {!}\, o\multimap o \, \, ) \multimap \,  {!}\, o \, \multimap \, o.
$$
As expected, the higher-order recursion scheme~$\mathcal{G}$ in~(\ref{equation/hors})
can be translated as a proof~$t_A$ of this formula~$A$ in linear logic extended with a fixpoint operator~$Y$.
An amusing and slightly puzzling observation is that the scheme~$\mathcal{G}$ can be alternatively
translated as a proof~$t_B$ of the formula~$B$ below:
$$
B \quad = \quad !\, (\, \, o \, \multimap \, o \, \multimap o \, \, )\multimap \, {!}\, (\, \, o\multimap o \, \, ) \multimap \,  {!}\, o \, \multimap \, o
$$
with the same underlying simply-typed $\lambda$-term with fixpoint operator~$Y$.
%
The difference between the terms~$t_A$ and~$t_B$ is not syntactic, but type-theoretic:
in the case of the term $t_A$, the type~$A$ indicates 
that each tree-constructor~$a$, $b$ and $c$ of the signature~$\Sigma$ 
is allowed to call its hypothesis as many times as desired:
$$
a \hspace{.5em} : \hspace{.7em} !\,o \, \multimap \,\, !\,o \, \multimap \, o
\quad\quad
b \hspace{.5em} : \hspace{.5em} !\,o \, \multimap \, o
\quad\quad
c \hspace{.5em} : \hspace{.5em} o
$$
whereas in the case of the term~$t_B$, the type~$B$ indicates
that each variable~$a,b,c$ calls each of its hypothesis exactly once:
$$
a \hspace{.3em} : \hspace{.3em} o \, \multimap \, o \, \multimap \, o
\quad\quad
b \hspace{.3em} : \hspace{.3em} o \, \multimap \, o
\quad\quad
c \hspace{.3em} : \hspace{.3em} o
$$
As a matter of fact, it appears that the proof~$t_B$ is the image
of the proof~$t_A$ along a canonical coercion of linear logic
$$\vdash \iota:A \multimap B.$$
%
%
The status of this program transformation~$\iota$ is difficult to understand
%
%
%
unless one recalls that linear logic is based on the existence
of a perfect duality between the programs
of a given type~$A$ and their environments or counter-programs
which are typed by the linear negation~$A^{\bot}$ of the original type~$A$.
%
%
Accordingly, since the two terms~$t_A$ and~$t_B=\iota\circ t_A$ are syntactically equal,
their difference should lie in the class of counter-programs of type~$A^{\bot}$ or~$B^{\bot}$
which are allowed to interact with them.
This idea takes its full flavour in the context of model-checking, when one realizes that every 
tree automaton~$\mathcal{A}$ on the signature~$\Sigma$ may be seen as a counter-program
whose purpose is indeed to interact with $t_A$ or $t_B$ 
in order to check whether a specific property of interest is satisfied by the infinite tree~$\valG$
generated by the recursion scheme~$\mathcal{G}$.
%
This leads to the tentative duality principle:
\begin{center}
\begin{tabular}{ccc}
\begin{tabular}{c}
higher-order\\
recursion schemes~$\mathcal{G}$
\end{tabular}
& \quad $\leftrightsquigarrow$ \quad \quad &
tree automata~$\mathcal{A}$
\end{tabular}
\end{center}
where a tree automaton~$\mathcal{A}$ on the signature~$\Sigma$
is thus seen as a counter-program of type~$A^{\bot}$ or~$B^{\bot}$
interacting with the higher-order recursion scheme~$\mathcal{G}$
seen as a program of type~$A$ or~$B$.
An apparent obstruction to this duality principle 
is that, in contrast to what happens with recursion schemes~$\mathcal{G}$,
it is in general impossible to translate a tree automaton~$\mathcal{A}$
as a proof of linear logic --- in particular because linear logic lacks non-determinism.
%
However, one neat way to resolve this matter and to extend linear logic
with non-determinism is to embed the logic in its relational semantics,
based on the monoidal category $Rel$ of sets and relations.
%
The relational semantics of linear logic is indeed entitled 
to be seen as a non-deterministic extension of linear logic
where every nondeterministic tree automaton 
$\mathcal{A}=\langle \Sigma,\,Q,\,\delta,q_0\rangle$ may be ``implemented''
by interpreting the base type~$o$ as the set~$Q$ of states of the automaton, 
and by interpreting each variable $a,b,c$ as the following relations
$$
a \hspace{.3em} : \hspace{.3em} Q \, \multimap \, Q \, \multimap \, Q
\quad\quad
b \hspace{.3em} : \hspace{.3em} Q \, \multimap \, Q
\quad\quad
c \hspace{.3em} : \hspace{.3em} Q
$$
deduced from the transition function~$\delta$ of the automaton:
$$\begin{array}{c}
a=\{(q_1,q_2,q) \in Q\times Q\times Q \,\, | \,\, (1,q_1)\wedge(2,q_2)\in\delta(q,a)\} 
\\
b=\{(q_1,q) \in Q\times Q \,\, | \,\, (1,q_1)\in\delta(q,b)\} 
\\
c=\{q \in Q \,\, | \,\, \delta(q,c)=true\} 
\end{array}$$
%
%
The nondeterministic tree automaton~$\mathcal{A}$ is then interpreted
as the counter-program $\mathcal{A}_B= {!}\,a\otimes {!}\,b\otimes {!}\,c \otimes d$ of type
$$
B^{\bot} \, =  \,\, \, !\,(Q \, \multimap \, Q \, \multimap \, Q) \,\, \otimes \,\, 
! \, (Q \, \multimap \, Q) \,\, \otimes \,\, !\, Q \,\, \otimes \,\, Q^{\bot}.
$$
obtained by tensoring the three relations $a,b,c$ lifted with by the exponential modality~$!$
together with the singleton $d=\{q_0\}$ consisting of the initial state of the automaton,
and understood as a counter-program of type~$Q^{\bot}$.
Note that by composition with the contraposite $\iota^{\bot}:B^{\bot}\multimap A^{\bot}$
of the coercion~$\iota$, one gets a counter-program $\mathcal{A}_A=\iota^{\bot}\circ\mathcal{A}_{B}$ of type
$$
A^{\bot} \, =  \,\, \, !\,( \, ! \, Q \, \multimap \, \, ! \, Q \, \multimap \, Q)
\,\, \otimes \,\, 
! \, (! \, Q \, \multimap \, Q)
\,\, \otimes \,\, 
!\, Q \,\, \otimes \,\, Q^{\bot}.
$$
Note also that when the type $o$ is interpreted as~$Q$ in the relational model, 
the counter-programs of type~$B^{\bot}$ of the form ${!}\,a\otimes {!}\,b\otimes {!}\,c\otimes d$ 
with $d=\{q_0\}$ correspond exactly to the non-deterministic tree automata on the signature~$\Sigma$
with set of states~$Q$ and initial state~$q_0$.
The difference between the two types~$A$ and~$B$ becomes very clear and meaningful
at this stage: shifting to the type~$A^{\bot}$ enables one to extend the class 
of nondeterministic tree automata to nondeterministic \emph{alternating} tree automata~$\mathcal{A}$
with typical transitions of the form
\begin{equation}\label{equation/alternating-transition}
\delta(q,a) \quad = \quad (1,q_1) \, \, \wedge \, \, (1,q_2)
\end{equation}
meaning that the tree automaton~$\mathcal{A}$
meeting the tree~$a(t_1,t_2)$ at state~$q$ explores 
the left subtree~$t_1$ \emph{twice} with states~$q_1$ and~$q_2$
and does not explore \emph{at all} the right subtree~$t_2$.
Such a transition~$\delta(q,a)$ is typically represented in the relational semantics
of linear logic by the singleton relation
\begin{equation}\label{equation/relational-transition}
a \, = \, \{\,\, (  \, \{\!|q_1,q_2|\!\} \, , \, \emptyset \, , \, q \, ) \,\, \}
 \,\, : \,\, {!}\,Q\multimap {!}\,Q\multimap Q
\end{equation}
where one uses the set~$!\,Q$ of finite multisets of~$Q$
to encode the transition~(\ref{equation/alternating-transition}) 
with the finite multiset~$\{\!|q_1,q_2|\!\}$ consisting of the two states~$q_1,q_2\in Q$
and the empty multiset~$\emptyset$ of states.
It should be stressed that a tree automaton~$\mathcal{A}$ admitting such an ``alternating'' transition~$\delta(q,a)$ 
\emph{cannot} be encoded as a counter-program of type~$B^{\bot}$ because
the transitions of the tree automaton~$\mathcal{A}$ are \emph{linear} in that type
and thus explore \emph{exactly} once each subtree~$t_1$ and $t_2$
of the tree~$a(t_1,t_2)$.
Summarizing the current discussion, we are entitled to consider that 
each linear type~$A^{\bot}$ and~$B^{\bot}$ reflects a specific class
of tree automata on the signature~$\Sigma$:
\begin{center}
\begin{tabular}{ccc}
$B^{\bot}$ & $\leftrightarrow$ &non-deterministic tree automata
\\
$A^{\bot}$ & $\leftrightarrow$ & non-deterministic \emph{alternating} tree automata
\end{tabular}
\end{center}
Accordingly, the purpose of the coercion~$\iota$ from $t_A$ to $t_B$ 
is to restrict the power of the class of alternating non-deterministic tree automata
allowed to explore the infinite tree~$\valG$
generated by the higher-order recursion scheme~$\mathcal{G}$
of signature~$\Sigma$.

\subsubsection*{\textbf{Description of the paper}}
The purpose of this paper is to show that this duality between recursion schemes
and tree automata underlies several of the recent developments in the field
of higher-order model-checking.
To that purpose, we start by establishing in \S \ref{section/type-theoretic-approach}
an equivalence between  the intersection type system introduced by Kobayashi
to describe infinitary coinductive proofs, and an infinitary variant of the traditional
relational semantics of linear logic developed in \cite{fossacs}.
This correspondence between an intersection type system and a relational semantics
of linear logic adapts to the field of higher-order model-checking ideas dating back
to Coppo, Dezani, Honsell and Longo \cite{coppo-et-al} and recently revisited by Terui \cite{terui}
and independently by de Carvalho \cite{carvalho} in order to establish complexity properties
of evaluation in the simply-typed $\lambda$-calculus.
It may be also seen as an account based on linear logic of the semantic approach
to higher-order model-checking developed by Aehlig in the early days of the field~\cite{aehlig}.
The main contribution of the paper is the observation developed in \S\ref{section/colors-as-a-parametric-comonad}
that this correspondence
between intersection type systems and the relational semantics of linear logic
extends to alternating parity games, and thus to the full hierarchy of the modal $\mu$-calculus.
This extension relies on the construction of a \emph{parametric comonad}
in the sense of \cite{parametric-comonad} defined as a family of modalities $\Box_{m}$
indexed by colors (or priorities) $m\in\mathbb{N}$, equipped with a series
of structural morphisms satisfying suitable coherence properties.
The resulting intersection type system provides a clean and conceptual explanation
for the type system designed by Kobayashi and Ong \cite{kobayashi-ong}
in order to accomodate the hierarchy of colors.
In particular, we show that a simpler but equivalent treatment of colors is possible.
Finally, we explain in \S\ref{section/colors-as-a-parametric-comonad}
in what sense the parametric comonad exhibited at the level of intersection types
corresponds to a traditional notion of exponential modality at the level of the relational semantics
of linear logic.
We obtain in this way a semantic reformulation of alternating parity games 
in an infinitary and colored variant of the relational semantics of linear logic.
In particular, just as for finitary tree automata, a state~$q\in Q$ of the alternating parity automaton
is accepted if and only if it is an element of the composite (in the relational semantics)
of the recursion scheme with the tree automaton.

\section{The type-theoretic approach to higher-order model-checking}
\label{section/type-theoretic-approach}
Developing an idea by Hosoya, Pierce and Vouillon \cite{hosoya-pierce-vouillon},
Kobayashi designed in \cite{kobayashi} a type-theoretic account of higher-order
model-checking in the particular case of an alternating tree automata (ATA)
testing for coinductive properties --- and thus \emph{without} parity conditions.
In this section, we briefly recall his terminology and results, 
and explain the hidden connection with relational semantics.

\subsection{Recursion schemes and simply-typed $\lambda$-terms with fixpoint}
Given a ranked alphabet $\Sigma$, we will consider in this paper
two kinds of $\Sigma$-labelled trees of finite or countable depth:
\begin{itemize}
\item \emph{ranked} trees, typically generated by higher-order recursion schemes,
in which the number of children of a node labelled with $f \in \Sigma$
is equal to the arity $\arity(f)$ of its label,
\item \emph{unranked} trees, typically used to describe run-trees of alternating automata,
and where the previous arity constraint is relaxed.
\end{itemize}
Given a base type $\bot$, we will consider the set $\setkinds$ of simple types, generated by the grammar
$$
\kappa\ \ ::=\ \bot \ \vert \ \kappa \rightarrow \kappa
$$
modulo associativity of the arrow to the right. Every simple type has a unique decomposition
$$
\kappa \ \ =\ \ \kappa_1 \rightarrow \cdots \rightarrow \kappa_n \rightarrow \bot
$$
where $n$ is the \emph{arity} of $\kappa$, denoted $\arity(\kappa)$.
The complexity of $\kappa$ is typically
measured by its \emph{order}, defined inductively by $\order(\kappa)\, =\,  0$ if $n\,=\,0$ and
$$
\order(\kappa) \ = \ 1 + \max(\order(\kappa_1),\ldots,\,\order(\kappa_n))
$$
In the sequel, following Kobayashi \cite{kobayashi}, we shall refer to simple types as \emph{kinds}, to prevent confusions with intersection types. We write $f\,::\,\kappa$ or $t\,::\,\kappa$ when a symbol $f$ or a term $t$ has kind $\kappa$.
The formalism of \emph{higher-order recursion schemes} (HORS)
on a given ranked signature~$\Sigma$ may be seen as equivalent 
to simply-typed $\lambda$-calculus with a recursion operator $Y$ 
and free variables $f\in \Sigma$ of order at most $1$.
Consequently, every free variable $f\in\Sigma$ of the $\lambda Y$-term
corresponding to the recursion scheme has kind
\begin{equation}\label{equation/arity-of-f}
\underbrace{\bot \ \rightarrow\ \cdots \ \rightarrow \ \bot}_{\arity(f)} \ \rightarrow \ \bot.
\end{equation}
where $\arity(f)$ denotes the arity of the terminal $f\in\Sigma$.
%
%
The normalization of the $\lambda Y$-term associated to the recursion scheme $\G$
produces a potentially infinite ranked tree, labelled by its free variables.
As we explained in the introduction, this translation of higher-order recursion schemes
into $\lambda Y$-terms may be seen as an instance of the Church encoding 
of ranked trees over the signature $\Sigma$.

%
%
In order to check whether a given monadic second-order formula holds
at the root of the infinite tree generated by a HORS, a traditional procedure
is to explore it using an \emph{alternating parity automaton} (APT).
Every exploration of the APT produces a run-tree labelled by the same signature, 
but unranked because of the alternating nature of the automaton.
The subclass of APT in which every state of the automaton is assigned color $0$
(the least coinductive priority) defines the class of \emph{alternating tree automata} (ATA),
which can test coinductive properties like safety, but cannot test inductive properties
like reachability.
The definitions of HORS and APT, as well as the correspondence 
between APT and monadic second-order logic, are recalled in the Appendix.
In the sequel, $S$ denotes the start symbol of a recursion scheme $\G$, 
$\mathcal{N}$ its set of non-terminals, and $\mathcal{R}(F)$ denotes 
for every non-terminal $F \in \mathcal{N}$ the simply-typed $\lambda$-term
it rewrites to as $F \rG \mathcal{R}(F)$ in the recursion scheme $\G$.
%
%
%


\subsection{From kinds to intersection types}
In his original work, Kobayashi reduces the study of coinductive properties
of higher-order recursion schemes to the definition of an intersection type system.
%
%
The general idea is that every transition of an alternating tree automaton
\begin{equation}\label{equation/transition-function}
\delta(q_0,\texttt{if}) = (2,q_0)\wedge (2,q_1)
\end{equation}
may be understood type-theoretically as a refinement of the simple type 
$$
\ift\,::\,\bot \rightarrow \bot \rightarrow \bot
$$
and reformulated as an \emph{intersection type} 
$$
\emptyset \rightarrow (q_0 \wedge q_1) \rightarrow q_0.
$$
This intersection type expresses the fact that, given any tree $T_1$ and a tree $T_2$ accepted
from both states $q_0$ and $q_1$, the composed tree $\ift \ T_1 \ T_2$ is accepted 
from the state~$q_0$.
%
%
%
%
Following this connection, Kobayashi defines for every HORS~$\G$ 
and every alternating tree automaton~$\APT$
\emph{without parity condition} a type system~$\Koba$ satisfying
the following property:
\begin{theorem}[Kobayashi \cite{kobayashi}]\label{th/kobayashi}
The sequent
$$
\vdash_{\G,\APT} \ S\,:\,q\,::\,\bot
$$
is provable in $\Koba$ if and only if there is a run-tree of $\APT$ over $\valG$ with initial state $q$.
\end{theorem}
Note that the intersection type system $\Koba$ is somewhat ad hoc since it depends
on~$\G$ and~$\APT$, in contrast to the approach developed in the present paper, 
based on a Church encoding of~$\G$ and~$\APT$ in a single intersection type system formulated
in \S\ref{section/colors-as-a-parametric-comonad} and \S\ref{section/colored-tensorial-logic}.

\subsection{Intersection types and the relational semantics of linear logic}
As a warm-up to the next two sections \S\ref{section/colors-as-a-parametric-comonad}
and \S\ref{section/colored-tensorial-logic},
and to the modal treatment of colors in alternating parity automata (APT),
we explain here in the simpler coinductive case, how to relate Kobayashi's 
intersection type system for alternating tree automata (ATA)
to an infinitary variant of the relational semantics of linear logic.
%
As already explained,
the Church encoding of a ranked tree over the signature $\Sigma\,=\,\{f_i\,:\,ar_i \, | \, i \in I\}$
defines a $\lambda Y$-term $t$ of simple type $\bot$ with free variables $f_i$ 
of kind~(\ref{equation/arity-of-f}) translated as the following formula of linear logic:
$$
f_i \quad : \quad \underbrace{!\,\bot \multimap \cdots \multimap\ !\,\bot}_{ar_i} \multimap \bot 
\quad\quad\quad \mbox{for $i\in I$.}$$
The $\lambda Y$-term $t$ itself is thus typed by the following sequent of linear logic:
$$
\cdots \quad , \quad f_i \quad : \quad !\, \Big( \hspace{.5em} \underbrace{!\,\bot \multimap \cdots \multimap\ !\,\bot}_{a_i} \multimap \bot \hspace{.5em} \Big) 
\quad , \quad \cdots \quad \vdash \quad t \quad : \quad   \bot 
$$
From this follows that its interpretation $\interp{t}$ in the relational semantics of linear logic
defines a subset of the following set of ``higher-order states''
$$
\interp{t} \ \subseteq\ \left[\!\!\!\left[\hspace{.5em}\bigotimes_{i\in I}\ !\, \left(\underbrace{!\,\bot \multimap \cdots \multimap\ !\,\bot}_{ar_i} \multimap \bot \right) \multimap \bot \hspace{.5em}\right]\!\!\!\right]
$$
where the return type $\bot$ is naturally interpreted as the set of states $\interp{\bot} = Q$
of the alternating tree automaton.
As explained in the introduction, the transition function $\delta$
of the alternating tree automaton $\APT$ is itself interpreted as a subset
$$
\interp{\delta} \quad \subseteq \quad \left[\!\!\!\left[ \hspace{.5em} \bigwith_{i\in I}\ \left(\underbrace{!\,\bot \multimap \cdots \multimap\ !\,\bot}_{ar_i} \multimap \bot \right) \hspace{.5em} \right]\!\!\!\right]
$$
which may be ``strengthened'' in the categorical sense as a subset
$$
\interp{\delta^{\dagger}} \quad \subseteq \quad \left[\!\!\!\left[ \hspace{.5em} \bigotimes_{i\in I}\hspace{.8em} ! \hspace{.5em} \left(\underbrace{!\,\bot \multimap \cdots \multimap\ !\,\bot}_{ar_i} \multimap \bot \right) \hspace{.5em} \right]\!\!\!\right]
$$
where we turn to our advantage the well-known isomorphism of linear logic:
$$
! \, ( \, A \, \& \, B \, ) \quad \cong \quad ! A \, \tensor \, ! B.
$$
As explained in the introduction, a first contribution of the article is to establish the following result
in the case of the traditional relational semantics of linear logic, extended here
with a fixpoint operator $Y$:
\begin{theorem}\label{th/itrs}
An alternating tree automaton $\mathcal{A}$ with a set of states $Q$
has a finite accepting run-tree with initial state $q_0$
over the possibly infinite tree generated by a $\lambda Y$-term $t$
if and only if there exists $u \in \interp{\delta^{\dagger}}$ 
such that $(u, q_0) \in \interp{t}$, where $\interp{\delta^{\dagger}}=\mathcal{M}_{fin}(\interp{\delta})$
denotes the set of finite multisets of elements of $\interp{\delta}$.
\end{theorem}
Another equivalent way to state the theorem is that the set of accepting states $q_0$
for a finite run-tree of the alternating tree automaton $\mathcal{A}$ is equal to 
the composition of $\interp{t}$ and of $\interp{\delta^{\dagger}}
$ in the relational semantics.
At this point, it appears that the only hurdle towards an extension of this theorem 
to the alternating tree automata with coinductive (rather than inductive)
acceptance condition is the \emph{finiteness} of multiplicities
in the traditional relational interpretation of the exponential modality.
For this reason, the authors developed in a companion paper~\cite{fossacs}
an \emph{infinitary} variant $\Relinfinitary$ of the relational model of linear logic, 
%
%
where the exponential modality noted there <<$\,\superbang$>> in order to distinguish it
from the traditional <<$\,!\,$>> transports every set $A$ (of cardinality required to be smaller than the reals) to the set
$$
\superbang A \ \ =\ \ \mathcal{M}_{count}(A)
$$
 of \emph{finite-or-countable} multisets of elements of $A$.
%
%
In this alternative relational semantics, there is a \emph{coinductive} fixpoint operator~$Y$
satisfying the equations of a Conway operator, and thus providing
an interpretation of the $\lambda Y$-calculus.
%
%
The infinitary interpretation of a $\lambda Y$-term is denoted $\interpinf{t}$
in order to distinguish it from the traditional finitary interpretation.
Note that the interpretation $\interpinf{\delta}=\interp{\delta}$ is unchanged,
and that its strengthening to $\interpinf{\delta^{\dagger}}$ reflects now
the infinitary principles of the model.
In particular, it is possible to detect whether a transition has been called 
a countable number of times.
This brings us to the second main contribution of this article, 
which is to adapt the previous theorem for finite accepting run-trees
to the general case of possibly infinite accepting run-trees:

\begin{theorem}\label{th/itrs2}
An alternating tree automaton $\mathcal{A}$ with a set of states $Q$
has a possibly infinite accepting run-tree with initial state $q_0$
over the possibly infinite tree generated by a $\lambda Y$-term~$t$
if and only if there exists $u \in \interpinf{\delta^{\dagger}}$ 
such that $(u, q_0) \in \interpinf{t}$, where $\interpinf{\delta^{\dagger}}=\mathcal{M}_{count}(\interpinf{\delta})$
denotes the set of finite-or-countable multisets of elements of $\interpinf{\delta}$.
\end{theorem}
This theorem should be understood as a purely semantic counterpart to Theorem \ref{th/kobayashi}.
The connection is provided by the foundational and elegant work by Bucciarelli and Ehrhard \cite{ill1,ill2}
on indexed linear logic, which establishes a nice correspondence between 
the elements of the relational semantics and a finitary variant of intersection types.
By shifting from finite to finite-or-countable multisets and intersection types,
we are able to recover here the discriminating power of general alternating tree automata.
In particular, the set of accepting states of the alternating tree automaton $\mathcal{A}$
is equal to the composition of $\interpinf{t}$ and of $\interpinf{\delta^{\dagger}}$ 
in our infinitary variant of the relational semantics.
We should mention however that there is a minor difference between our semantic result
and the original Theorem \ref{th/kobayashi}, related to the fact that Kobayashi chose
to work in a type system where intersection is understood as an idempotent operation.
This choice is motivated in his work by the desire to keep the type system finitary,
and thus to obtain decidability results.
We prefer to work here with an infinitary relational semantics, corresponding
to an infinitary and non-idempotent variant of Kobayashi's intersection type system.
The reason is that shifting from an infinitary to an idempotent intersection type system
corresponds from a semantic point of view to shifting from an infinitary relational semantics 
to its extensional collapse.
Ehrhard \cite{ehrhard-collapse} has recently established that the extensional collapse of the relational semantics
is provided by a lattice model, where the formulas of linear logic are interpreted
as partially ordered sets.
This means that the corresponding intersection type systems should include
a subtyping relation, as well-understood for instance by Terui in \cite{terui}.
This subtyping relation is not mentioned in the original work by Kobayashi \cite{kobayashi}
nor in the later work by Kobayashi and Ong \cite{kobayashi-ong} and although their final
result is certainly valid, this omission has lead to much confusion.
%

%

\section{A type-theoretic account of alternating parity automata}
\label{section/colors-as-a-parametric-comonad}
\subsection{Colored intersection types}
\label{section/colored-intersection-types}

After designing in~\cite{kobayashi} the type-theoretic approach to alternating tree automata
recalled in the previous section, Kobayashi carried on in this direction
and generalized it with Ong~\cite{kobayashi-ong}
to the larger class of alternating \emph{parity} automata.
The basic idea of this work is to incorporate coloring annotations in the intersection types, 
in order to reflect in the type system the parity conditions of the tree automata.
Suppose for instance that a binary terminal $a\in \Sigma$ induces a transition
$\delta(q,a)=(1,q_1)\wedge (2,q_2)$ in an alternating parity tree automaton
with coloring function $\Omega:Q\to \mathbb{N}$.
In that case, the terminal $a$ is assigned in~\cite{kobayashi-ong} the intersection type
\begin{equation}\label{equation/terminal-a-typed-in-ko}
a \quad : \quad \ \,  (q_1,m_1) \rightarrow (q_2,m_2) \rightarrow q
\end{equation}
where $m_1=max(\Omega(q_1),\Omega(q))$ and $m_2=max(\Omega(q_2),\Omega(q))$
are colors indicating to the type system the colors of the states $q, q_1, q_2$ of the parity tree automaton.
In order to prepare the later development of paper, we find useful
to simplify the colored intersection type system originally formulated by Kobayashi and Ong,
and to stress at the same time the modal nature of colors (or priorities) in higher-order model-checking.
%
%
So, given a set of states $Q$ and a coloring function $\Omega\,:\,Q \rightarrow \mathbb{N}$,
we define the set of colors
$$
Col \ \ = \ \ \left\{\,\Omega(q)\ \vert\ q \in Q\,\right\} \ \uplus \ \left\{\,\epsilon\,\right\}
$$
which contains the colors used by $\Omega$, together with an additional color~$\epsilon$
which will play the role of neutral element.
The intersection types are then generated by the grammar
\begin{center}
\begin{tabular}{rclcr}
$\theta$ &$ \ ::=\ $ & $q \ \ \vert\ 
 \ \tau \flechedimplication \theta $ & $\ $ & $(q \in Q)$\\
$\tau$& $ \ ::=\ $ & $\bigwedge_{i \in I}\,\KObox_{m_i}\ \theta_i$ & &$(I\ \text{finite}, \,m_i \in Col)$\\
\end{tabular}
\end{center}
%
%
%
The refinement relation between intersection types and kinds
is defined by the inductive rules below:
$$
\begin{tabular}{ccccc}
$
\AxiomC{$q\in Q$}
\UnaryInfC{$q::\bot$}
\DisplayProof
$
&
\hspace{1.5cm}
&
$
\AxiomC{$\tau :: \kappa_1$}
\AxiomC{$\theta :: \kappa_2$}
\BinaryInfC{$\tau \rightarrow \theta :: \kappa_1 \rightarrow \kappa_2$}
\DisplayProof
$
&
\hspace{1.5cm}
$
\AxiomC{$\forall i \in I\ \,\,\, \theta_i :: \kappa$}
\UnaryInfC{$\bigwedge_{i \in I} \KObox_{m_i}\theta_i :: \kappa$}
\DisplayProof
$
\end{tabular}
$$
\noindent
Note that the color modality acts on intersection types and contexts by
$$
\begin{tabular}{ccc}
$
\KObox_m \, \left(\,\bigwedge_{i \in I}\KObox_{m_i}\,\theta_i \,\right) \quad = \quad
\bigwedge_{i \in I}\KObox_{\operatorname{max}(m,m_i)}\,\theta_i
$
&
\hspace{0.8cm}
&
$
\KObox_m \left( \, x: \tau \, , \, \Delta \, \right)
\quad = \quad
x: \KObox_m\, \tau \,\, , \,\, \KObox_m \, \Delta
$
\end{tabular}
$$

\noindent
Note also that the neutral color $\epsilon$ is only introduced here to allow a uniform definition of types and contexts.
It does not affect the coloring of types, and should be understood
as the \emph{absence} of a coloring annotation.
From this, one obtains an intersection type system $\KOnew$
parametrized by the alternating tree automaton~$\mathcal{A}$,
whose rules are given in Figure \ref{figure/KOnew}.
Here we use the Hebrew letter $\mbox{\tsadi}$ which should be read ``tsadi''.
The resulting type system $\KOnew$ enables us 
to type the rewriting rules of a higher-order recursion scheme
\begin{equation}\label{eq/KO-typing}
\Delta \ \vdash\ \mathcal{R}(F)\,:\,\sigma\,::\,\kappa
\end{equation}
where the non-terminals occurring in the $\lambda$-term $\mathcal{R}(F)$
appear as variables in the context~$\Delta$ of the typing judgement.
On the other hand, the intersection type system $\KOnew$ does not include
a fixpoint operator~$Y$ and for that reason does not accomodate recursion.
%

\begin{figure*}[t!]
\small
$$
\AxiomC{$$}
\LeftLabel{Axiom $\quad \quad$}
\RightLabel{$\quad \quad \left( x \in \mathcal{V} \cup \mathcal{N} \right)$}
\UnaryInfC{$x \, : \, \bigwedge_{\{i\}}\,\KObox_{\epsilon}\ \theta_i \,::\, \kappa \hspace{.5em} \vdash \hspace{.5em} x\, : \, \theta_i \,::\, \kappa$}
\DisplayProof
$$

$$
\AxiomC{$\{ \, (i,q_{ij})  \, \, | \,  \, 1\leq i\leq n , 1\leq j\leq k_i\}$ \,  satisfies \, $\delta_{A}(q,a)$\, }
\LeftLabel{$\delta$ \quad}
\RightLabel{\quad $a \in \Sigma$}
\UnaryInfC{$\emptyset \vdash a \, : \, \bigwedge_{j=1}^{k_1}\ \KObox_{\Omega(q_{1j})}\ q_{1j} \, \flechedimplication \, \dots \, \flechedimplication \, \bigwedge_{j=1}^{k_n}\ \KObox_{\Omega(q_{nj)}}\ q_{nj} \flechedimplication q\, ::\, \bot \rightarrow \cdots \rightarrow \bot \rightarrow \bot$}
\DisplayProof
$$

$$
\AxiomC{$\Delta\vdash t : \left(\KObox_{m_1}\ \theta_1\ \wedge \dots\wedge \KObox_{m_k}\ \theta_k \right) \flechedimplication \theta \, :: \, \kappa \flechedimplication \kappa'$}
\AxiomC{$\quad \Delta_1\vdash u \,  :  \, \theta_1 \, ::\, \kappa \quad \cdots \quad
\Delta_k\vdash u \,  :  \, \theta_k ::\, \kappa$}
\LeftLabel{App \quad \quad}
\BinaryInfC{$\Delta \, + \, \KObox_{m_1} \Delta_1\, + \, \dots \, + \, \KObox_{m_k} \Delta_k \hspace{.5em} \vdash \hspace{.5em} t \, u \, : \, \theta ::\, \kappa'$}
\DisplayProof
$$

$$
\AxiomC{$\Delta \, , \, x \, : \, \bigwedge_{i\in I} \, \KObox_{m_i}\ \theta_i \, :: \, \kappa \hspace{.5em} \vdash \hspace{.5em} t \, : \, \theta \, ::\, \kappa'$}
\AxiomC{$\quad\quad I \, \subseteq \, J$}
\LeftLabel{$\lambda$ \quad \quad}
\BinaryInfC{$\Delta \hspace{.5em} \vdash \hspace{.5em} \lambda \, x \, . \, t \, : \, \left( \bigwedge_{j\in J} \, \KObox_{m_j}\ \theta_j  \right)  \flechedimplication \theta ::\, \kappa \flechedimplication \kappa'$}
\DisplayProof
$$
\normalsize
\caption{The type system $\KOnew$ associated to the alternating parity tree automaton~$\APT$.}
\label{figure/KOnew}
\end{figure*}

\label{section/adamic-edenic}
%

\subsection{Interpretation of recursion}
%
%
%
In order to accomodate recursion in the intersection type system~$\KOnew$,
we need to extend it with a rule $fix$ whose purpose is to expand
the non-terminals $F\in\mathcal{N}$ of the recursion scheme~$\G$
in order to obtain possibly infinitary derivation trees.
%
%
So, given a higher-order recursion scheme $\G$ and an alternating parity automaton $\mathcal{A}$,
we define the intersection type system $\KOYnew$ as $\KOnew$ where we add
the recursion rule

$$
\AxiomC{$\Gamma \vdash \mathcal{R}(F)\,:\, \theta\,::\,\kappa$}
\LeftLabel{$fix \quad \quad\quad$}
\RightLabel{$\quad\quad \quad \operatorname{dom}(\Gamma) \subseteq \mathcal{N}$}
\UnaryInfC{$F \,: \, \KObox_{\epsilon}\ \theta \,::\, \kappa \,\, \vdash \,\, F \,: \, \theta \,::\, \kappa $}
\DisplayProof
$$
and at the same time restrict the Axiom rule to variables $x\in \mathcal{V}$, 
and in particular do not allow the Axiom rule to be applied on non-terminals any more.
\noindent
An important aspect of the resulting intersection type system $\KOYnew$
is that its derivation trees may be of countable depth. 
As in Kobayashi's original type system, this infinitary nature of the intersection type system
enables one to reflect the existence of infinitary runs in the alternating parity automaton~$\mathcal{A}$.
In order to articulate the parity condition of the automaton with the typing derivations,
a color is assigned to each node of the derivation tree, in the following way:
\begin{itemize}
\item the node $\Delta_i\vdash u \,  :  \, \theta_i \, ::\, \kappa $ is assigned the color $m_i$
in every Application rule
$$
\AxiomC{$\Delta\vdash t : \left(\KObox_{m_1}\ \theta_1\ \wedge \dots\wedge \KObox_{m_k}\ \theta_k \right) \flechedimplication \theta \, :: \, \kappa \flechedimplication \kappa'$}
\AxiomC{$\quad \cdots \quad \Delta_i\vdash u \,  :  \, \theta_i \, ::\, \kappa \quad \cdots$}
\BinaryInfC{$\Delta \, + \, \KObox_{m_1} \Delta_1\, + \, \dots \, + \, \KObox_{m_k} \Delta_k \hspace{.5em} \vdash \hspace{.5em} t \, u \, : \, \theta ::\, \kappa'$}
\DisplayProof
$$
of the derivation tree,
\item all the other nodes of the derivation tree are assigned the neutral color $\epsilon$,
which means in some sense that they are not colored by the typing system.
\end{itemize}
A nice aspect of our approach compared to the original formulation in \cite{kobayashi-ong}
is that the parity condition traditionally applied to the alternating parity automaton~$\mathcal{A}$
extends to a very simple parity condition on the derivation trees of $\KOYnew$.
Indeed, the color of an infinite branch of a given derivation tree can be defined as
\begin{itemize}
\item the neutral color $\epsilon$ if no other color $m \in Col$ occurs infinitely often
in the branch, 
\item otherwise, the maximal non-neutral color $m\in Col \setminus \left\{\epsilon\right\}$
seen infinitely often.
\end{itemize}
Then, an infinite branch of the derivation tree is declared \emph{winning}
precisely when its color is an even integer (and in particular different from the neutral color).
A winning derivation tree is then defined as a derivation tree
whose infinite branches are all winning in the sense just explained.

\subsection{Soundness and completeness}
Once the notion of infinite winning derivation tree explicated, as we have just done
in the previous section, there remains to relate this winning condition to
the acceptance condition of alternating parity automata.
To that purpose, and for the sake of the presentation,
we choose to restrict ourself to \emph{productive} recursion schemes,
as it is also done in \cite{kobayashi-ong}.
Note that it is only a very mild restriction, since every recursion scheme $\G$ can be
transformed in to a productive recursion scheme $\G'$ which outputs a special leaf 
symbol $\Omega$ whenever the B\"ohm  evaluation of the original scheme $\G$
would have infinitely looped.
The following theorem establishes a soundness and completeness theorem
which relates the winning condition on the infinite derivation trees of $\KOYnew$
to the parity acceptance condition of the automaton~$\mathcal{A}$ 
during its exploration of the infinite tree $\valG$ generated by the recursion scheme~$\G$:


\begin{theorem}[soundness and completeness]
\label{th/KOY-typing}
Suppose given a productive recursion scheme $\G$
and an alternating parity automaton $\APT$.
There exists a winning run-tree of $\APT$ over $\valG$ with initial state $q$ if and only if the sequent
\begin{equation}\label{equation/sequent-at-starting-point}
S\,:\,\KObox_{\epsilon}\ q\,::\,\bot \ \vdash \ S\,:\,q\,::\,\bot
\end{equation}
has a winning derivation tree in the type system $\KOYnew$.
\end{theorem}
There are several ways to establish the theorem.
One possible way is to establish an equivalence with the original soundness
and completeness by Kobayashi and Ong~\cite{kobayashi-ong}.
One should be careful however that the original proof in~\cite{kobayashi-ong} 
was incomplete, and has been corrected in the (unpublished) journal version of the paper.
In order to establish the equivalence, one shows that the existence of a winning derivation tree 
of the sequent (\ref{equation/sequent-at-starting-point}) in the intersection type system $\KOYnew$
is equivalent to the existence of a winning strategy for Eve in the parity game
defined in \cite{kobayashi-ong}.
Another more direct proof is possible, based on the reformulation by Haddad \cite{these-axel}
of Kobayashi and Ong's treatment of infinitary (rather than simply finitary) sequences
of rewrites on higher-order recursion schemes.
It is in particular important to observe that the infinitary nature of computations
requires to extend the usual soundness and completeness arguments based
on B\"ohm trees, finite rewriting sequences and continuity.
This point was apparently forgotten in \cite{kobayashi-ong} and corrected in the journal version
of the paper.

\section{An indexed tensorial logic with colors}
\label{section/colored-tensorial-logic}
The notation $\KObox_{m}\, \theta$ is used in our intersection type system $\KOYnew$
as a way to stress the modal nature of colors,
and it replaces for the better the notation $(\theta,m)$ used by Kobayashi and Ong
in \cite{kobayashi-ong}.
As we will see, the discovery of the modal nature of colors is fundamental,
and is not just a matter of using the appropriate notation.
In particular, it enables us to simplify both technically
and conceptually the original intersection type system in \cite{kobayashi-ong}.
By way of illustration, the original intersection type (\ref{equation/terminal-a-typed-in-ko})
of the binary terminal $a\in\Sigma$ considered in \S\ref{section/colored-intersection-types}
is replaced by 
the simpler intersection type:
\begin{equation}\label{equation/terminal-a-typed-in-the-modal-system}
a \quad : \quad \ \,  \KObox_{n_1}\ q_1 \rightarrow \KObox_{n_2}\ q_2 \rightarrow\, q
\end{equation}
where $n_1=\Omega(q_1)$ and $n_2=\Omega(q_2)$.
Interestingly, the color of the state $q$ is not mentioned in the type anymore.
The reason is that this alternative account of colors achieved
in our type system is not just ``simpler'' than the original one: 
it also reveals a deep and somewhat unexpected connection with linear logic, 
since as we will see, this ``disparition'' of the color $\Omega(q)$ in
(\ref{equation/terminal-a-typed-in-the-modal-system}) is related
to the well-known linear decomposition $A\Rightarrow B = {!}A \multimap B$ 
of the intuitionistic implication in linear logic.
One essential difference however is that the exponential modality <<\,!\,>> of linear logic
is replaced by a family of modal boxes $\Omega(m)$ which formally defines
what Melli\`es calls a \emph{parametric comonad}
in \cite{mellies:parametric-continuation-monad}\cite{mellies:functorial-boxes}.

This key observation enables us to translate the intersection type system $\KOYnew$ 
into an infinitary variant of linear logic equipped with a family of color modalities
noted $\Box_m$ for $m\in\mathbb{N}$.
A nice feature of the translation is that it transports the intersection type system $\KOYnew$ 
which depends on $\mathcal{G}$ and $\mathcal{A}$ into an intersection type system
which does not depend on them anymore --- although it still depends 
on the set $Q$ of states of the automaton.
The infinitary variant of linear logic which we use for the translation is
\begin{itemize}
\item \emph{indexed} in the sense of Bucciarelli and Ehrhard~\cite{ill1,ill2}.
In particular, the finite or countable intersection types $\bigwedge_{i \in I} \theta_i$ 
of $\KOYnew$ are translated as finite or countable indexed families $[ \theta_i\ |\ i\in I]$
of formulas of the logic,
\item \emph{tensorial} in the sense of Melli\`es \cite{tensorial_logic,mellies_gs_string,mellies:parametric-continuation-monad}. In this specific case, every negated
formula of the logic is negated with respect to a specific state $q\in Q$ 
of the automaton, and is thus of the form $\sigma\multimap q$,
which may be alternatively written as $\lnot_q \, \sigma$ or even as $\stackrel{q}{\lnot} \sigma$.
\end{itemize}
In this way, one obtains an indexed and colored variant of tensorial logic,
called $\textrm{LT}(Q)$ in the sequel, and whose formulas are inductively
generated by the following grammar:
$$
A, B \quad ::= \quad
1 
\quad \vert \quad 
A\otimes B
\quad \vert \quad 
\lnot_q \, A
\quad \vert \quad 
\square_m\, A  
\quad \vert \quad 
[\, A_j\ \vert\ j \in J\, ]
  \quad \quad (m \in Col,\ q \in Q)
$$
As already mentioned, following the philosophy in~\cite{ill2},
the finite or countable indexed set $[\sigma_j\ \vert\  j \in J]$
internalizes the intersection operator of $\KOYnew$ in our indexed tensorial logic,
see our companion paper \cite{itrs} for details.
Importantly, the resulting indexed logic $TL(Q)$ can be used as an intersection type system 
refining the simply-typed $\lambda$-calculus in just the same way as $\KOYnew$,
see the Appendix~\label{app/colored-tensorial-types}.
In particular, the fact that $\square$ defines a parametric monoidal comonad
in the logic means that the sequents
\begin{center}
\begin{tabular}{ccc}
$\square_{\epsilon}\ A$ & $\ \ \vdash\ \ $ & $A$\\
$\square_{\operatorname{max}(m_1,\,m_2)}\ A$ & $\ \ \vdash\ \ $ & $\square_{m_1}\ \square_{m_2}\ A$\\
$\square_m\ A \ \otimes\ \square_m\ B$ & $\ \ \vdash\ \ $ & $\square_m\ (A \ \otimes\ B)$\\
\end{tabular}
\end{center}
are provable for all colors $m,m_1,m_2\in\mathbb{N}$, 
and all formulas~$A,B$.
%
%
In order to deal with recursion schemes,
 we admit derivation trees with finite or countable depth in the logical system $TL(Q)$.
The nodes of the derivation trees of $TL(Q)$ are then colored in the following way:
%
\begin{itemize}
\item every node $\Gamma\vdash M \,  :  \, A \, ::\, \kappa $
in a Right introduction of the modality $\Box_m$ :
$$
\AxiomC{$\Gamma \vdash \, M \, : \, A \, :: \, \kappa$}
\RightLabel{\quad Right $\square_m$} 
\UnaryInfC{$\square_m\, \Gamma \, \vdash \, M \, : \, \square_m\, A \, :: \, \kappa$}
\DisplayProof
$$
is assigned the color~$m$ of the modality,
\item all the other nodes of the derivation tree are assigned the neutral color $\epsilon$.
\end{itemize}
The winning condition on an infinite derivation tree of $TL(Q)$ is then directly adapted
from the similar condition in $\KOYnew$.
%
%
Thanks to this condition, we are ready to state a useful correspondence
theorem 
between $\KOYnew$ and $TL(Q)$
for any (productive) recursion scheme $\G$. 
Suppose that for each $F \in \mathcal{N}$ of kind $\kappa(F)$ of the recursion scheme~$\G$,
we introduce a new free variable $freeze(F)$ of kind $\kappa(F) \rightarrow \kappa(F)$ ; 
that we replace each $\lambda$-term $\mathcal{R}(F)$ by its $\beta\eta$-long normal form ;
and finally, that we substitute each occurrence of $F$ appearing 
in any $\beta\eta$-long normal form $\mathcal{R}(G)$ of the recursion scheme~$\mathcal{G}$
with the $\lambda$-term~$freeze(F)\ F$ of the same kind $\kappa(F)$.
This transformation induces a context-free grammar of <<\,blocks>> consisting
of the $\beta\eta$-long $\mathcal{R}(G)$'s,
which generates an infinite $\lambda$-term in $\beta\eta$-long normal form, 
noted~$term(\G)$, with free variables of the form $freeze(F)$.
%
Moreover, this infinite $\lambda$-term $term(\G)$ is coinductively typed
in the simply-typed $\lambda$-calculus by the typing judgment:
\begin{equation}\label{equation/typing-judgment}
\dots\quad  ,\quad freeze(F):\kappa(F)\to\kappa(F) \quad ,\quad \dots \quad \vdash\quad term(\G)\quad : \quad \bot
\end{equation}
where $F$ runs over all the non-terminals $F\in\mathcal{N}$ of the higher-order recursion scheme~$\G$.
At this point, we are ready to recast our Theorem~\ref{th/KOY-typing} in the proof-theoretic
language of indexed tensorial logic:
\begin{theorem}\label{th/colored-tensorial-logic}
There exists a winning derivation tree in $\KOYnew$ of the sequent
\begin{equation}\label{eq/sequent-KOY}
S\,:\,\KObox_{\epsilon}\ q_0\,::\,\bot \ \vdash \ S\,:\,q_0\,::\,\bot
\end{equation}
if and only if there exists a winning derivation tree in $TL(Q)$ of a sequent
\begin{equation}\label{eq/sequent-tensoriel}
\Gamma \ \vdash \ \ term(\G)\,:\,q_0\,::\,\bot
\end{equation}
refining the typing judgment~(\ref{equation/typing-judgment}).
\end{theorem}

\section{Putting all together: relational semantics of linear logic
and higher-model model-checking}
\label{section/colored-rel}
Once the connection between higher-order model-checking and 
indexed tensorial logic established in \S\ref{section/colored-tensorial-logic},
there remains to exhibit the associated relational semantics of linear logic,
following the ideas of Bucciarelli and Ehrhard~\cite{ill1,ill2}.
%
%
%
%
%
%
%
This trail leads us to an infinitary and colored variant of the usual relational semantics of linear logic,
developed in our companion paper~\cite{fossacs}.
%
The key observation guiding the construction is that the functor 
$$
\square\quad : \quad A \quad \mapsto \quad Col \times A \quad : \quad Rel \quad \to \quad Rel
$$
equipped with the coercion maps
\begin{center}
\begin{small}
\begin{tabular}{ccrcl}
$\left\{\left(\left(\left(m,a\right),\left(m,b\right)\right),\left(m,\left(a,b\right)\right) \right)\ \vert \ a \in A,\, b \in B,\, m\in Col\right\}$ & $:$ & $\square\, A \, \otimes\, \square\, B$ & $\ \rightarrow \ $ & $\square\, \left(\,A \, \otimes \, B\,\right)$\\
$\left\{\,\left(\star,\left(m,\star\right)\right)\ \vert\ m \in Col \,\right\}$ & $:$ &$1$ & $\ \rightarrow \ $ & $\square\, 1 $\\
$\left\{\,\left(\,\left(\,\operatorname{max}(m_1,\,m_2),\,a\right),\,\left(\,m_1,\,\left(m_2,\,a\right)\,\right)\,\right) \ \vert \ a \in A\,\right\}$ & $:$ &$\square\ A$  & $\ \rightarrow \ $ & $\square\ \square\ A$ \\
$\left\{\,\left(\,\left(\,\epsilon,\,a\right),\,a\right) \ \vert\ a \in A\,\right\}$ & $:$ &$\square\ A $ & $\ \rightarrow \ $ & $A$ \\
\end{tabular}
\end{small}
\end{center}
defines a lax monoidal comonad $\Box:Rel\to Rel$ on the category $Rel$ of sets and relations.
Moreover, the comonad distributes (or better: commutes) with the exponential modality $\superbang$, 
in such a way that these two comonads compose into a new exponential modality
of linear logic $\colorbang$ defined by the equation $\colorbang A = \superbang \, \Box \, A$. 
A Conway operator $Y$ can be then defined in order to reflect in the relational semantics
the definition of the winning condition on the infinite derivations of $TL(Q)$.
%
%
This fixpoint operator can be seen as a combination of the inductive and coinductive fixpoints
of the model, where the color of an input indicates whether the fixpoint operator
should be defined inductively (when the color is odd or neutral) or coinductively (when the color is even).
The relational interpretation of a $\lambda Y$-term $t$ 
in this infinitary model is denoted as $\interpcoltest{t}$. 
%
The interpretation $\interpcoltest{\delta}$ of the transition function $\delta$
is defined similarly as for $\interpinf{\delta}$, except that the color information
is incorporated in the semantics following the comonadic principles
underlying the translation~(\ref{equation/terminal-a-typed-in-the-modal-system}) 
in \S\ref{section/colored-tensorial-logic}.
Typically, the transition (\ref{equation/transition-function}) is interpreted 
in the colored relational semantics as
$$
([\, ],([(\Omega(q_0),q_0),(\Omega(q_1),q_1)],q_0)) \quad \in \quad \interpcoltest{\delta}
$$
The last contribution of this paper, which underlies our companion paper \cite{fossacs}
but is not stated there, establishes a clean correspondence between the relational semantics
of a higher-order recursion scheme $\mathcal{G}$ (seen below as a $\lambda\,Y$-term~$t_{\mathcal{G}}$) and the exploration
of the associated ranked tree $\valG$ by an alternating \emph{parity} automaton~$\mathcal{A}$:
\begin{theorem}\label{th/relational-model-checking}
An alternating \emph{parity} tree automaton $\mathcal{A}$ with a set of states $Q$
 has a \emph{winning} run-tree with initial state $q_0$
over the ranked tree $\valG$ generated by the $\lambda Y$-term~$t_{\mathcal{G}}$
if and only if there exists $u \in \interpcol{\delta^{\dagger}}$ 
such that $(u, q_0) \in \interpcol{t_{\mathcal{G}}}$, where $\interpcol{\delta^{\dagger}}=\mathcal{M}_{count}(Col \times \interpcol{\delta})$
denotes the set of finite-or-countable colored multisets of elements of $\interpcol{\delta}$.
\end{theorem}



\section{Related works}
\label{section/related-works}
The field of higher-order model-checking was to a large extent started at the turn of the century
by Knapik, Niwinski, Urzyczyn, who established that for every $n\geq 0$, $\Sigma$-labelled trees
generated by order-$n$ safe recursion schemes are exactly those that are generated by
order-$n$ pushdown automata, and further, that they have decidable MSO theories.
The safety condition was relaxed a few years later by Ong,
who established the MSO decidability for general order-$n$ recursion schemes,
using ideas imported from game semantics.
Unfortunately, Ong's proof was intricate and somewhat difficult to understand.
%
Much work was thus devoted to establish the decidability result by other means.
Besides the type-theoretic approach initiated by Kobayashi~\cite{kobayashi,kobayashi-ong},
Hague, Murawski, Ong, Serre~\cite{cpda} developed an automata-theoretic approach based
on the translation of the higher-order recursion scheme~$\mathcal{G}$ into a collapsible pushdown automaton (CPDA),
which led the four authors to another proof of MSO decidability for order-$n$ recursion schemes.
A clarifying connection was then made by Salvati and Walukiewicz
between this translation of higher-order recursion schemes into CPDAs
and the traditional evaluation mechanism of the environment Krivine machine~\cite{krivine-hors}.
Following this discovery, Salvati and Walukiewicz are currently developing
a semantic approach to higher-order model checking, based on the interpretation
in finite models of the $\lambda$-calculus 
with fixpoint operators, see~\cite{salvati,salvati2} for details.
%
%
%
The idea of connecting linear logic to automata theory is a longstanding dream
which has been nurtured by a number of important contributions.
%
%
%
Among them, we would like to mention the clever work by Terui~\cite{terui}
who developed a semantic and type-theoretic approach 
based on linear logic, intersection types and automata theory
in order to characterize the complexity of evaluation
in the simply-typed $\lambda$-calculus.
%
%
In a different but related line of work, explicitly inspired by Bucciarelli and Ehrhard's indexed linear logic~\cite{ill1,ill2},
de Carvalho~\cite{carvalho} establishes an interesting correspondence between intersection types
and the length of evaluation in a Krivine machine.

%

%


\section{Conclusions and perspectives}

\label{section/conclusion}
The purpose of the present paper is to connect higher-order model-checking
to a series of advanced ideas in contemporary semantics,
like linear logic and its relational semantics, indexed linear logic,
distributive laws and parametric comonads.
All these ingredients meet and combine surprisingly well.
%
The approach reveals in particular that the traditional treatment of inductive-coinductive reasoning
based on colors (or priorities) is secretly based on the same comonadic principles
as the exponential modality of linear logic.
Besides the conceptual promises offered by these connections, we would like to conclude the paper
by mentioning that this stream of ideas leads us to an alternative and purely semantic
proof of MSO decidability for higher-order recursion schemes, after~\cite{ong, kobayashi-ong,cpda}.
The basic idea is to replace the infinitary colored relational semantics constructed in \S\ref{section/colored-rel}
by a finitary variant based on the prime-algebraic lattice semantics of linear logic.
From a type-theoretic point of view, this lattice semantics corresponds 
to an intersection type system with subtyping for linear logic recently formulated by Terui~\cite{terui}.
We have shown in a companion paper \cite{icalp} how to recover the MSO decidability result
for order-$n$ recursion schemes by adapting to this finitary semantics of linear logic
the constructions performed here for its relational semantics.
One interesting feature of the resulting model of the $\lambda\,Y$-calculus 
is that a morphism $D\to E$ in the Kleisli category consists in a continuous function
$$
\vspace{-.2em}
f \quad : \quad \underbrace{D\times \dots \times D}_{n} \quad \morph{} \quad E
\vspace{-.1em}
$$
where $n$ is the number of colors considered in the semantics ; 
and that the fixpoint $Yf$ of a morphism $f:D\to D$ is defined in that case by the alternating formula
$$
Y f \quad = \quad \nu x_{n} \, . \,  \mu x_{n-1} \, \dots \, \nu x_2 \, . \, \mu x_1\, . \,  \nu x_0 . f(x_0,\dots, x_n)
$$
where we suppose (without loss of generality) that $n$ is even, where 
$\mu$ and $\nu$ denote the least and greatest fixpoint operators, respectively.
We believe that the apparition of this simple formula and the fact that it defines 
a Conway operator~$Y$ and thus a model of the $\lambda\,Y$-calculus
is a key contribution to the construction of a semantic and purely 
compositional account of higher-order model-checking.

\bibliographystyle{plain}
\bibliography{tensorial-logic-with-colors}

\newpage

\appendix


\section{Logical specification and automata theory}
\label{section/logic-automata}

\subsection{Monadic second-order logic and modal $\mu$-calculus}

The purpose of higher-order model-checking
is to abstract the behavior of a functional program with recursion as a tree approximating the set of its potential executions,
and then to specify a logical property to check over this tree. 
The tradition in higher-order model-checking is to consider monadic second-order logic,
a well-balanced choice between expressivity -- it contains most other usual logics over trees --
and complexity: the satisfiability of a formula is decidable for infinite structures of interest -- as infinite trees (Rabin 1969).
Higher-order verification has a different approach: the question is whether a \emph{given tree}
satisfies the formula -- or whether an equivalent automaton accepts it.
A first step towards this automata model for MSO~is
\begin{theorem}[Janin-Walukiewicz 1996]
%
MSO is equi-expressive to modal $\mu$-calculus over trees.
\end{theorem}
\noindent
where modal $\mu$-calculus formulae are defined by
$$
\phi\ ::=\ X\ \vert\ \underbar{$f$} \ \vert \ \phi \vee \phi \ \vert \ \phi \wedge \phi \ \vert \ \square \, \phi \ \vert \ \diamond_i \phi\ \vert \ \mu X.\, \phi\ \vert \ \nu X.\, \phi
$$
for $f \in \Sigma$. Given a ranked tree, the semantics of a formula is the set of nodes where it holds.
The predicate $\underbar{f}$ is true on $f$-labelled nodes, $\square\ \phi$ is true on nodes
whose succesors all satisfy $\phi$, $\diamond_i\ \phi$ is true on nodes whose $i^{th}$ succesor
satisfies $\phi$, and the $\mu$ and $\nu$ are two fixpoints operators which can be understood in two different manners.
Semantically, they are dual operators,
%
$\mu$ and $\nu$ being respectively the least and
greatest fixpoint on the semantics of formulae.

Given a $\Sigma$-labelled ranked tree whose set of nodes is $N$, whose branching structure
is given by a finite family of successor functions $succ_i\,:\,N \rightarrow N$,
and whose labelling is described by a function $label\,:\,N \rightarrow \Sigma$,
the semantics of a closed modal $\mu$-calculus formula $\phi$ is defined as $\semantics{\phi}_{\emptyset}$
where $\emptyset$ denotes the unique function $\emptyset \rightarrow N$, and
for a function $\mathcal{V}\,:\,Var \rightarrow N$ and a modal $\mu$-calculus formula
$\psi$, the semantics $\semantics{\psi}_{\mathcal{V}}$ are defined inductively:

\begin{itemize}
\item $\semantics{a}_{\mathcal{V}} \, =\, \{ n \in N \ \vert \ label(n)\ =\ a\}$
\item $\semantics{X}_{\mathcal{V}} \, =\, \mathcal{V}(X)$
\item $\semantics{\neg \phi}_{\mathcal{V}} \, =\, N \setminus \semantics{\phi}_{\mathcal{V}}$
\item $\semantics{\phi \vee \psi}_{\mathcal{V}} \, =\,\semantics{\phi}_{\mathcal{V}} \cup \semantics{\psi}_{\mathcal{V}}$
\item $\semantics{\diamond_i\, \phi}_{\mathcal{V}}\, =\, \{n \in N \ \vert \ ar(n) \geq i \mbox{ and } succ_i(n) \in \semantics{\phi}_{\mathcal{V}}\}$
\item $\semantics{\mu X.\,\phi(X)}_{\mathcal{V}} \, =\,\bigcap\ \{M \subseteq N \ \vert\ \semantics{\phi(X)}_{\mathcal{V}[X \leftarrow M]} \subseteq M  \} $
\end{itemize}

where $\mathcal{V}[X \leftarrow M]$ coincides with $\mathcal{V}$ except on $X$ which it maps to $M$.
The semantics of $\wedge$, $\square$ and $\nu$ are defined using de Morgan duality.

Another understanding of $\mu$ and $\nu$ is syntactic and closer to automata theory:
both allow the unfolding of formulae
$$
\mu X.\,\phi[X] \ \rightarrow_{\mu} \phi[\mu X.\,\phi[X]] \mbox{  and  } \nu X.\,\phi[X] \ \rightarrow_{\nu} \phi[\nu X.\,\phi[X]]
$$
but $\rightarrow_{\mu}$ may only be expanded finitely, while $\rightarrow_{\nu}$ is unconstrained. The semantics of a formula may then be understood as the set of positions from which it admits unfoldings which are logically true and which do not use $\rightarrow_{\mu}$ infinitely.

\subsection{Alternating parity tree automata}

From this syntactic interpretation of fixpoints over formulae, we can define a class of automata corresponding
to modal $\mu$-calculus, namely \emph{alternating parity automata} (APT), whose purpose is to synchronise
the unraveling of formulas with symbols of the tree.
These automata are top-down tree automata,
with two additional features:
\begin{itemize}
\item \emph{alternation}: they have the power to duplicate or drop subtrees,
and to run with a different state on every copy,
\item and \emph{parity conditions}: since run-trees are infinitary by nature, these automata
discriminate \emph{a posteriori} the run-tree unfolding $\rightarrow_{\mu}$ infinitely.
\end{itemize}

The transition function takes values in positive Boolean formulae over couples of states and directions, its generic shape being
\begin{equation}\label{equation/apt-transition}
\delta(q,\,a) \ \ =\ \ \bigvee_{i \in I} \ \bigwedge_{j \in J_i} (d_{i,j}, \, q_{i,j})
\end{equation}
which consists of a non-deterministic choice of $i$ followed by the execution of $\vert J_i \vert$ copies of the automaton,
each on the successor in direction $d_{i,j}$ of the current node, with state $q_{i,j}$.
%
%
%

When for every $i \in I$ and every direction $d$ there is a unique
$j$ such that $d_{i,j} = d$, we recover the usual notion of \emph{non-deterministic} parity automaton.
States of an APT may be understood as subformulae of the formula of interest, so that some correspond
to subformulae $\mu X.\,\phi$ and others to subformulae $\nu X.\,\phi$. To exclude infinite unfoldings of $\mu$,
every state $q$ is given a \emph{color} $\Omega(q) \in \mathbb{N}$. States in the immediate scope of a $\mu$
receive an odd color, and the others an even one. If $q$ corresponds to a subformula of $q'$, then the coloring will
satisfy $\Omega(q) \leq \Omega(q')$. The construction of $\Omega$ is such that the greatest color among the ones
seen infinitely often in an infinite branch informs the automaton about which fixpoint operator was unfolded infinitely along it.

A branch of a run-tree is \emph{winning} when the greatest color seen infinitely often along it is even.
A run-tree is declared \emph{winning} when all its infinite branches are.
Every modal $\mu$-calculus formula $\phi$ can be translated to an APT $\mathcal{A}_{\phi}$ such that
\begin{theorem}[Emerson-Jutla 1991]
Given a $\Sigma$-labelled ranked tree $T$, $\phi$ holds at the root of $T$ if and only if $\mathcal{A}_{\phi}$ has a
winning run-tree over $T$.
\end{theorem}

\subsection{An interactive interpretation of APT}

Recall that a \emph{parity game} is a graph in which each vertex $v \in \mathcal{V}$ belongs to a player: Eve or Adam.
It can be understood as a game where a token moves from vertex to vertex, starting from the initial one,
and taking on each vertex an outgoing edge chosen by the player who controls it. The resulting interaction is called a \emph{play},
and a maximal play is finite if and only if it ends on a vertex without outgoing edges.
There is a coloring function $\Omega\,:\,\mathcal{V}\rightarrow \mathbb{N}$, and the winning condition over infinite
plays is defined just as for infinite branches of run-trees. For finite maximal plays, the player controlling the last vertex loses.

A \emph{strategy} for a player is a map from the set of plays ending with a node he controls to $\mathcal{V}$.
It indicates the player which move he should take during a play.
It is \emph{positional} if it can be recovered from a function $\mathcal{V} \rightarrow \mathcal{V}$.
The strategy is \emph{winning} (resp. \emph{colorblind}) if every maximal play (resp \emph{finite} maximal play)
in which it is followed by the player is winning for him.

\begin{theorem}[Martin 1975]\label{th/parity-games}
Parity games enjoy \emph{positional determinacy}: given an initial vertex, one of the players has a winning positional
strategy from it. It is computable when the game is finite.
\end{theorem}

%
%

The execution of an APT over a tree $T$ may then be understood as a parity game in which Eve constructs a run-tree
by playing the non-deterministic choice of the transition function (\ref{equation/apt-transition}): she selects $i$,
while Adam chooses a direction to explore by picking $j\in J_i$. A play is thus an exploration of a branch -- controlled by Adam --
of a run-tree built by Eve. Then Eve has a winning strategy from the root (and the initial state)
if and only if she can build a run-tree in which Adam can not find a branch that violates the parity condition
or is rejected by the automaton: she has a winning strategy if and only if $\APT$ has a winning run-tree over $T$.
%


\section{Higher-order recursion schemes}
\label{section/hors}

Functional programs are a challenge for verification, as they feature higher-order recursion.
Higher-order recursion schemes (HORS) provide an abstract model of functional programs which precisely
focuses on the complex program flow induced by this recursive power.
HORS produce \emph{trees} abstracting the set of executions of programs.
They notably do not allow the evaluation of conditionals nor the treatment of references.

Consider a signature $\Sigma$, a set of variables $\mathcal{V}$, and a set of \emph{non-terminals} $\mathcal{N}$.
The function $\kind$ is extended to $\mathcal{V} \uplus \mathcal{N}$ with a simple type for each variable and
non-terminal. A HORS is the data of an \emph{axiom} $S \in \mathcal{N}$
%
of simple type $\bot$ and of a function $\mathcal{R}$ mapping each non-terminal $\mathcal{N}$ to a closed term
\begin{equation}\label{eq/hors-rule}
\mathcal{R}(F) \ =\ \lambda x_1.\,\ldots \lambda x_n.\, t
\end{equation}
of simple type $\kind(F)$, and such that each of the $x_i$ is in $\mathcal{V}$ and that $t$ is a term
without abstractions.

The \emph{order} of $\mathcal{G}$  is
$\operatorname{max}\left(\left\{order(\kind(F))\ \vert\ F \in \mathcal{N} \right\}\right)$.
We define inductively the rewriting relation $\rG$ over terms by:
\begin{itemize}
\item $F\,t_1\, \cdots \, t_n \ \rG \ t[x_i := t_i]$ if $\mathcal{R}(F)\,=\,\lambda x_1 \cdots \lambda x_n .\, t$,
\item if $s \rG t$ then $s\,u \rG t\,u$ and $u\,s \rG u\,t$.
\end{itemize}
The \emph{value tree} $\valG$ of the scheme, when it exists,
%
is defined as the limit tree obtained by this infinite rewriting process starting from $S$, and is a $\Sigma$-labelled ranked tree.

We define the term $\termG$  as the one obtained by considering $\mathcal{R}$ as a regular grammar.
It is the infinite term corresponding to $\mathcal{G}$; its $\beta$-reduction computes $\valG$.

\begin{example} Given $\Sigma\,=\,\{\,\texttt{if}\,:\,2,\,\texttt{data}\,:\,1,\,\texttt{Nil}\,:\,0,\}$, consider the recursion scheme
$$
\begin{tabular}{rcl}
$\texttt{S}$ & $\quad = \quad $ & $\texttt{L Nil}$\\
$\texttt{L}$ & $\quad = \quad $ & $\lambda x.\,\texttt{if } x\ (\texttt{L }( \texttt{data } x\ ) $\\
\end{tabular}
$$

Its value tree is depicted in Figure \ref{fig/value-tree}. Even though the scheme is very simple,
this is not a regular tree: it has infinitely many different subtrees. A consequence is that
the application of Theorem \ref{th/parity-games} over $\valG$ does not suffice to decide the 
existence of a winning run-tree over it.
The effect of the transitions
$$
\delta(q_0,\texttt{if}) = (2,q_0)\wedge (2,q_1)
\mbox{ and }
\delta(q_1,\texttt{if})=(1,q_1)\wedge (2,q_0)
$$
is depicted in Figure \ref{fig/run-tree}.
\end{example}

\begin{figure*}
\small
\begin{minipage}{.5\textwidth}
$$
\begin{tikzpicture}
\tikzset{level distance=24pt}
\Tree [.$\ift$ $\Nil$ [.$\ift$ [.$\data$ $\Nil$ ] [.$\ift$ [.$\data$ [.$\data$ $\Nil$ ] ] $\vdots$ ] ] ]
\end{tikzpicture}
$$
\caption{An order-1 value tree.}
\label{fig/value-tree}
\end{minipage}
\begin{minipage}{.5\textwidth}
$$
\begin{tikzpicture}
\tikzset{level distance=24pt}
\Tree [.$\ift\ \ q_0$ [.$\ift\ \ q_0$ [.$\ift\ \ q_0$ $\vdots$ ] [.$\ift\ \ q_1$ $\vdots$ ] ] [.$\ift\ \ q_1$ [.$\data\ \ q_1$ $\vdots$ ] [.$\ift\ \ q_0$ $\vdots$ ] ] ]
\end{tikzpicture}
$$
\caption{An APT run-tree.}
\label{fig/run-tree}
\end{minipage}
\end{figure*}

%
%

\section{Connection with the Kobayashi-Ong approach}
\label{app/KO}

\subsection{The Kobayashi-Ong type system}
\label{app/KOsys}

Consider a coloring function $\Omega\,:\,Q\rightarrow \mathbb{N}$,
we extend it to intersection types by setting $\Omega(\tau \flechedimplication \sigma) = \Omega(\sigma)$.
The original type system of Kobayashi and Ong is recasted in Figure \ref{figure/kobayashi-ong}.
Note that every rule of a recursion scheme admits
a finite number of colored intersection typings (\ref{eq/KO-typing}), where the contexts consist of refined
typings of the non-terminals occuring in $\mathcal{R}(F)$. In a context $\Delta$, a non-terminal $G$ typically occurs as
\begin{equation}\label{eq/non-terminal-typing}
G\,:\,\bigwedge_{i \in I}\,\KObox_{m_i}\ \theta_i\,::\,\kind(G)
\end{equation}

\begin{figure}[t]
\small
$$
\AxiomC{$$}
\LeftLabel{Axiom $\quad \quad$}
\RightLabel{$\quad \quad \left( x \in \mathcal{V} \cup \mathcal{N} \right)$}
\UnaryInfC{$x \, : \, \bigwedge_{\{i\}}\,\KObox_{\Omega(\theta_i)}\ \theta_i \,::\, \kappa \hspace{.5em} \vdash \hspace{.5em} x\, : \, \theta_i \,::\, \kappa$}
\DisplayProof
$$

$$
\AxiomC{$\{ \, (i,q_{ij})  \, \, | \,  \, 1\leq i\leq n , 1\leq j\leq k_i\}$ \,  satisfies \, $\delta_{A}(q,a)$\, }
\LeftLabel{$\delta$ \quad}
\RightLabel{\quad for $a \in \Sigma$ and $m_{ij}=\operatorname{max}(\Omega(q_{ij}),\Omega(q))$}
\UnaryInfC{$\emptyset \vdash a \, : \, \bigwedge_{j=1}^{k_1} \KObox_{m_{1j}}\ q_{1j} \, \flechedimplication \, \dots \, \flechedimplication \, \bigwedge_{j=1}^{k_n}\KObox_{m_{nj}}\ q_{nj} \flechedimplication q\, ::\, \bot \rightarrow \cdots \rightarrow \bot \rightarrow \bot$}
\DisplayProof
$$

$$
\AxiomC{$\Delta\vdash t : \left(\KObox_{m_1}\ \theta_1\ \wedge \dots\wedge \KObox_{m_k}\ \theta_k \right) \flechedimplication \theta \, :: \, \kappa \flechedimplication \kappa'$}
\AxiomC{$\quad \Delta_1\vdash u \,  :  \, \theta_1 \, ::\, \kappa \quad \cdots \quad
\Delta_k\vdash u \,  :  \, \theta_k ::\, \kappa$}
\LeftLabel{App \quad \quad}
\BinaryInfC{$\Delta \, + \, \KObox_{m_1} \Delta_1\, + \, \dots \, + \, \KObox_{m_k} \Delta_k \hspace{.5em} \vdash \hspace{.5em} t \, u \, : \, \theta ::\, \kappa'$}
\DisplayProof
$$

$$
\AxiomC{$\Delta \, , \, x \, : \, \bigwedge_{i\in I} \, \KObox_{m_i}\ \theta_i \, :: \, \kappa \hspace{.5em} \vdash \hspace{.5em} t \, : \, \theta \, ::\, \kappa'$}
\AxiomC{$\quad\quad I \, \subseteq \, J$}
\LeftLabel{$\lambda$ \quad \quad}
\BinaryInfC{$\Delta \hspace{.5em} \vdash \hspace{.5em} \lambda \, x \, . \, t \, : \, \left( \bigwedge_{j\in J} \, \KObox_{m_j}\ \theta_j  \right)  \flechedimplication \theta ::\, \kappa \flechedimplication \kappa'$}
\DisplayProof
$$
\normalsize
\caption{The Kobayashi-Ong type system $\KO$ associated to the alternating parity tree automaton~$\APT$}
\label{figure/kobayashi-ong}
\end{figure}

\subsection{Recursion as a parity game}
Fixing a recursion scheme $\G$ and an APT $\APT$, we obtain a finite set of typings (\ref{eq/KO-typing}) for each rewrite rule.
In order to account for recursion, Kobayashi and Ong introduce the finite parity game $\Adamic$, in which Adam incrementally tries
to disprove Eve's typing by picking non-terminals to unfold.

More specifically, Eve's vertices correspond to colored typings for non-terminals, while Adam's vertices are typing contexts.
There is an edge from a typing $F\msp:\msp\KObox_{m}\ \theta\msp ::\msp\kappa$ to a context $\Delta$ if and only the sequent
\begin{equation}\label{eq/adamic-sequent}
\Delta \vdash \mathcal{R}(F)\,:\,\theta\,::\,\kappa
\end{equation}
is provable in the colored type system $\KO$, and there is an edge from a context $\Delta$ to a typing $G\msp:\msp\KObox_{m}\ \theta\msp::\msp\kappa$
if and only if $G$ occurs in $\Delta$ with this refined type -- that is, if $G$ occurs in $\Delta$ as in (\ref{eq/non-terminal-typing}),
if there is $i \in I$ such that $m\msp=\msp m_i$ and $\theta\msp =\msp \theta_i$. Note that the resulting game is finite,
due to the idempotency of the intersection operator.
Vertices $F\msp:\msp\KObox_{m}\ \theta\msp ::\msp\kappa$  receive color $m$; other vertices receive the neutral color $\epsilon$.

A \emph{play} is winning for Adam if and only if it ends on a node $F\msp:\msp\KObox_{m}\ \theta\msp ::\msp\kappa$ from
which Eve has no move - that is, if she made a typing assumption she can not prove - or if it is infinite and such that Adam could choose
infinitely often to expand a non-terminal of maximal odd color.
%
Therefore, Eve has a winning strategy in this game if and only if she can ensure the existence of a winning
sequence of typings along every possible branch of reductions in the scheme, leading to
\begin{theorem}[Kobayashi-Ong 2009 \cite{kobayashi-ong}]\label{th/kobayashi-ong}
Eve has a winning strategy in the parity game $\Adamic$ iff the alternating \emph{parity} tree automaton $\APT$ has a winning run-tree over $\valG$.
\end{theorem}

\subsection{From $\Adamic$ to $\Edenic$}

Despite its intuitive connection with type theory, the game $\Adamic$ does not describe the on-the-fly
construction of a branch of a typing proof, for two essential reasons:
\begin{itemize}
\item Eve does not provide a \emph{witness} of the typing proof of the sequent (\ref{eq/adamic-sequent})
which builds $\Delta$, so that proofs can not be extracted from plays, and that no distinction
is made between different derivations with the same conclusion,
\item and Adam does not play an \emph{occurence} of a non-terminal in $\mathcal{R}(F)$,
but a \emph{typing} which could, due to idempotency, correspond to several Axiom leaves of
a derivation tree.
\end{itemize}
In order to understand the game $\Adamic$ from a purely type-theoretic point of view,
we introduce the parity game $\Edenic$, which only differs on these two points:
\begin{itemize}
\item Eve plays \emph{typing proofs} of sequents (\ref{eq/adamic-sequent})
in addition to the context $\Delta$ they build,
\item and Adam plays \emph{occurences} of non-terminals appearing in the term
$\mathcal{R}(F)$ -- or, equivalently, picks an Axiom leaf introducing a non-terminal
in the proof $\pi$ provided by Eve at the previous turn.
\end{itemize}
Note that the resulting game is bigger, yet finite. We prove the following correspondence:

\begin{proposition}\label{prop/adamic-edenic}
The parity games $\Adamic$ and $\Edenic$ are equivalent, in the sense that a player
has a winning strategy in a game if and only if he does in the other.
\end{proposition}

Note that the collapse of strategies of $\Edenic$ to strategies of $\Adamic$ relies on a uniformization
property which is reminiscent of the proof of the positional determinacy of parity games: from a winning strategy for Eve
in $\Edenic$, one can extract such a strategy in which, given a colored type, any occurence of a non-terminal
with this type will be mapped to the same typing proof.

\subsection{Colored typings}

From the parity game $\Edenic$, we can easily define a corresponding type system similar in the spirit to $\KOYnew$,
but using Kobayashi and Ong's original coloring policy.
Consider the system $\KOY$ obtained from $\KO$ by restricting the Axiom rule to
variables $x\in \mathcal{V}$, and by adding the $\fixrule$ rule
$$
\AxiomC{$\Gamma \vdash \mathcal{R}(F)\,:\, \theta\,::\,\kappa$}
\LeftLabel{$fix \quad$}
\RightLabel{$\quad \operatorname{dom}(\Gamma) \subseteq \mathcal{N}$}
\UnaryInfC{$F \,: \, \KObox_{\Omega(\theta)}\ \theta \,::\, \kappa \,\, \vdash \,\, F \,: \, \theta \,::\, \kappa $}
\DisplayProof
$$
As in $\KOYnew$, derivations of infinite depth are allowed in $\KOY$.
Colorblind strategies $\sigma$ for Eve in $\Edenic$ are easily translated as proofs $\KOYtranslation{\sigma}$ in $\KOY$,
by incrementally plugging to every $fix$ rule the finite proof-tree she answers in $\sigma$,
starting from the unfolding of the axiom $S$ of the scheme
$$
\AxiomC{$\vdots$}
\UnaryInfC{$\Gamma \vdash \mathcal{R}(S)\,:\, q_0\,::\,\bot$}
\LeftLabel{$fix \quad$}
\UnaryInfC{$S \,: \, \KObox_{\Omega(q_0)}\ q_0 \,::\, \bot \,\, \vdash \,\, S \,: \, q_0 \,::\, \bot $}
\DisplayProof
$$
Of course, the inverse process can be defined as well, leading to the unique definition for every
derivation tree $\pi$ in $\KOY$ of a colorblind strategy $\Edenictranslation{\pi}$ for Eve in $\Edenic$.
Adam's r\^ole in $\Edenic$ is also transported in $\KOY$: its strategies leading to infinite interactions with $\sigma$
are in one-to-one bijection with the infinite branches of $\KOYtranslation{\sigma}$. 
Strategies resulting in finite interactions are not quite branches, but paths leading to an instance of a
$fix$ rule expanding a term without non-terminals.
%
%

In order to account for the parity condition in $\Edenic$, we color the $fix$ rules of the tree
in the same manner: every $fix$ rule expands a non-terminal occuring as (\ref{eq/non-terminal-typing})
in the context introduced by the immediately preceding instance of $fix$; it receives the color
$m_i$ corresponding to the refined type $\theta_i$ of the occurence to expand.
The first $fix$ rule, which expands $S$, receives color $0$.
Now the usual parity condition for trees is incorporated to $\KOY$ derivation
trees: \emph{winning} derivation trees are those whose infinite branches
all satisfy the parity condition.
\begin{theorem}\label{th/KOY}
\begin{itemize}
\item A colorblind strategy $\sigma$ for Eve is winning in $\Edenic$ if and only if
$\KOYtranslation{\sigma}$ is a winning derivation tree of $\KOY$.
\item A derivation tree $\pi$ of $\KOY$ is winning if and only if
Eve's colorblind strategy $\Edenictranslation{\pi}$ is winning in $\Edenic$.
\end{itemize}
\end{theorem}

\subsection{Proof of Theorem~\ref{th/KOY-typing}}

Theorem \ref{th/KOY-typing} is central in this article, as it discloses the comonadic behaviour of the coloring annotation
of the alternating parity automaton. It admits at least three proofs:
\begin{itemize}
\item one proof consists in a minor and at the same time clarifying alteration of the proof given by Kobayashi and Ong
in the unpublished journal version of their original article, 
\item another more direct proof is based on the equivalence between the game $\Edenic$
and of the game $\Edenicnew$ obtained by playing typing derivations of the $\KOnew$ type system
instead than in $\KO$, 
\item the authors are currently writing a third proof formulated in a purely proof-theoretic language
and based on standard techniques in linear logic and semantics, as well as on the proof
by Kobayashi and Ong in the unpublished journal version of their original article~\cite{kobayashi-ong}
and on the proof by Haddad~\cite{these-axel}.
\end{itemize}\ \\

\noindent
\textbf{Adapting Kobayashi and Ong's original proof.} We start by briefly explaining how the original proof of soundness
given in Kobayashi and Ong's unpublished journal version of~\cite{kobayashi-ong} may be very naturally adapted to the comonadic color policy formulated in our paper.
The key idea is to change the definition of the color $\Omega(C[\,]_q)$ of a context in the following way:
\begin{itemize}
\item  if $C[\,]_q\ = \ [\,]_q$, then $\Omega(C[\,]_q)\ =\ \epsilon$,
\item  if $C[\,]_q\ = \ \langle\,\alpha,\,q'\,\rangle\ T_1 \, \cdots \, T_{i-1}\ C'[\,]_q \ T_{i+1} \, \cdots\, T_n$, then
\begin{itemize}
\item if $C'[\,]_q\ =\ [\,]_q$, then $\Omega(C[\,]_q)\ =\ \epsilon$,
\item if $C'[\,]_{q}\ =\ \langle\,\beta,\,q''\,\rangle\ T'_1 \, \cdots \, T'_{j-1}\ C''[\,]_q \ T'_{j+1} \, \cdots\, T'_m$, then   $\Omega(C[\,]_q)\ =\ \operatorname{max}(\Omega(q''),\,\Omega(C'[\,]_{q}))$.
\end{itemize}
\end{itemize}

The definition is written here in the style of Kobayashi and Ong's proof. However,
in order to understand its true content, one should observe that every tree-context $C[\,]_q$ is either trivial or
has a  ``return state'' $q'$ defined as the state of the tree automaton labelling its root
$\langle\,\alpha,\,q'\,\rangle$.
The color of the trivial tree-context $[\,]_q$ is itself trivial (that is, equal to the neutral color $\epsilon$) while the color of a
non-trivial tree-context $C[\,]_q$ is equal to the maximal color encountered from its return state $q'$ (not included)
to its hole (not included). This reflects the comonadic nature of our coloring policy, as disclosed in \S\ref{section/colored-tensorial-logic}.
If we write $C[\,]_q^{q'}$ for a non-trivial tree-context with return type $q'$, we thus have that the color of a context
$$
\langle\,\alpha,\,q''\,\rangle\ T_1 \, \cdots \, T_{i-1}\ C[\,]_q^{q'} \ T_{i+1} \, \cdots\, T_n
$$
is equal to $\operatorname{max}\left(\Omega\left(q'\right),\,\Omega\left(C[\,]_{q}^{q'}\right)\right)$.

\noindent
To summarize, the general philosophy of tensorial logic is that the color $c\, =\, \Omega(q)$ which labels the state $q$ 
in Kobayashi and Ong's approach labels in our case the comonadic box $\square_{c}$ which encloses the state $q$.
The definition of Kobayashi and Ong's function $\Lambda$ should be adapted accordingly: in the case (i) of the definition of the
rewriting relation $\rhd$ in Section~4.1 of their unpublished journal version of~\cite{kobayashi-ong},
$\Lambda\{ l \mapsto \Omega(q)\}$
must be changed to $\Lambda\{ l \mapsto \epsilon\}$ in order to reflect the fact that the tree context is trivial in that case.
These revised definitions of $\Omega$ and $\Lambda$ clarify the secretely comonadic principles underlying the
proofs of Lemma 4.4 and 4.5 of Kobayashi and Ong's unpublished journal article, which for that reason can be easily
adapted to our new, comonadic coloring policy. The remaining part of the proof of soundness (in particular Lemma 4.8)
works exactly in the same way in our setting as in Kobayashi and Ong's original formulation. This provides a proof
of soundness for the system $\KOYnew$.

\ \\

\noindent
\textbf{Another proof directly based on parity games.} An alternative and simple way to prove
Theorem \ref{th/KOY-typing} (soundness and completeness) is to return to the original
framework of parity games, and in particular to the equivalence
between $\Adamic$ and $\Edenic$ established in Proposition~\ref{prop/adamic-edenic}.
Remember indeed that the parity game $\Edenic$ is closely related to the proof-system $\KO$.
We thus introduce a minor variant of this parity game called $\Edenicnew$
adapting $\Edenic$ to the proof-system $\KOnew$, which plays for $\KO$ the same role
as the system $\KOYnew$ for $\KOY$.
%
%
It is easy to recast Theorem~\ref{th/KOY} in
this setting, and to show that there is a correspondence between Eve's colorblind strategies in the parity game $\Edenicnew$
and the derivation trees in the type system $\KOYnew$. Moreover, this correspondence maps winning strategies to
winning derivations, and conversely. We shall now prove the following equivalence:

\begin{proposition}\label{prop/edenic-edenicnew}
If $\G$ is a \emph{productive} higher-order recursion scheme, then the parity games $\Edenic$ and $\Edenicnew$ are equivalent, in the sense that a player
has a winning strategy in one of these games if and only if he has a winning strategy in every of these games.
\end{proposition}


%

\smallbreak

\noindent
We have already established the equivalence between $\Adamic$ and $\Edenic$ in Proposition~\ref{prop/adamic-edenic}.
In order to prove Proposition \ref{prop/edenic-edenicnew}, we observe that there exists a one-to-one correspondence
between $\KO$ and $\KOnew$ derivation trees.
This implies that there exists a one-to-one correspondence
between Eve's colorblind strategies in $\Edenic$ and in $\Edenicnew$.
The same is true for Adam: his strategies in both games are identical.

\smallbreak

\noindent
In order to establish that the parity games $\Edenic$ and in $\Edenicnew$ are equivalent,
there remains to prove that this correspondence preserves the color of infinite branches.
Notice that the productivity of the recursion scheme $\G$ excludes the existence of an infinite branch containing,
after a finite prefix, only non-terminals and bound variables on head positions: 
indeed, their head reduction would lead to $\bot$, which is forbidden by productivity.
A consequence is that every infinite branch of a \emph{productive} recursion scheme $\G$
visits infinitely often a term in head position in an Application rule, 
so that every such infinite branch is assigned countably many non-neutral colors in $\Edenicnew$.
Of course, between two such visits to non-neutral colors, the neutral color $\epsilon$ is played in $\Edenicnew$,
whereas the last visited color is repeated in the original game $\Edenic$. 
However, since such repetitions are always of finite length, the maximal color
visited infinitely often in an infinite play of the strategy for Eve (resp. for Adam) in $\Edenic$
is precisely the same as the color visited infinitely often in the corresponding strategy
for Eve (resp. for Adam) in $\Edenicnew$.
As a consequence, the correspondence between $\Edenic$ and $\Edenicnew$ preserves
the colors of infinite plays and thus transports a winning strategy for Eve (resp. for Adam)
in $\Edenic$ to a winning strategy for Eve (resp. for Adam) in $\Edenicnew$ and conversely.
This completes the proof of Proposition~\ref{prop/edenic-edenicnew}.

Together with the equivalence between $\Adamic$ and $\Edenic$ established 
in Proposition~\ref{prop/adamic-edenic}, this equivalence induces Theorem~\ref{th/KOY-typing},
at least for a productive recursion scheme $\G$.
The soundness and completeness theorem follows indeed for the corresponding theorem
established by Kobayashi and Ong (Theorem~\ref{th/kobayashi-ong}) for the parity game~$\Adamic$.
The conceptual novelty of course of Theorem~\ref{th/KOY-typing} is that it reveals the comonadic nature of colors
in higher-order model-checking, and a promising connection of this field with linear logic.

\section{Typing terms with colored tensorial logic}
\label{app/colored-tensorial-types}

The indexed and colored variant $TL(Q)$ of tensorial logic with colors 
is introduced and discussed in \S\ref{section/colored-tensorial-logic}.
The main logical rules of the system are formulated below in Figure~\ref{fig/tensorial-calculus}.
In order to simplify their presentation, we choose to write
$$
\neg_q\ (A_1,\,\cdots,\,A_n)
$$
for the formula of indexed tensorial logic :
$$
\neg_q\ (A_1\,\otimes\,\cdots\,\otimes\,A_n) \quad  = \quad 
A_1\,\multimap \, \cdots\,\multimap \, A_n \, \multimap \, q.
$$
Similarly, and for the sake of uniformity, we choose to write 
$$
\neg\ \, (\kappa_1,\,\cdots,\,\kappa_n)
$$
for the type (or kind) of simply-typed $\lambda$-calculus:
$$
\kappa_1\,\rightarrow \,\cdots\,\rightarrow\,\kappa_n\rightarrow \bot.
$$
Note that, for the sake of simplicity, we prefer to keep implicit
the index $I,J$ or $K$ appearing on the side of each sequent
of the logical rules below.
The reader interested in the precise treatment of such indexes
will find the detailed treatment in the work by Bucciarelli and Ehrhard~\cite{ill1,ill2}
as well as in our companion paper \cite{itrs}.

\begin{figure}
\begin{center}

$$
\AxiomC{$q \in Q$}
\LeftLabel{Axiom \quad} 
\UnaryInfC{$x\,:\, q\, ::\, \bot\ \vdash\ x\, :\, q\, ::\, \bot$}
\DisplayProof
$$

\vspace{0.5cm}

\begin{tabular}{llr}

\AxiomC{$\Gamma,\,x\,:\,A\,::\,\kappa\ \vdash\ M\,:\,B\, :: \,\kappa'$}
\LeftLabel{Left $\square$\quad} 
\UnaryInfC{$\Gamma,\,x\,:\,\square_{\epsilon}\ A\,::\,\kappa\ \vdash\ M\,:\,B \,::\, \kappa'$}
\DisplayProof

&&

\AxiomC{$\Gamma\ \vdash\ M\,:\,A\, ::\, \kappa$}
\RightLabel{\quad Right $\square_m$} 
\UnaryInfC{$\square_m\ \Gamma\ \vdash\ M\,:\, \square_m\ A \, ::\, \kappa$}
\DisplayProof
\\
\end{tabular}

\vspace{0.5cm}

\begin{tabular}{llr}
\AxiomC{$\Gamma ,\, x\,:\,A\, ::\, \kappa\ \vdash\ M\, :\, B \, ::\, \kappa' $}
\LeftLabel{Dereliction \quad} 
\UnaryInfC{$\Gamma ,\, x\,:\,[A] \,::\, \kappa\ \vdash\ M\, : B\,::\, \kappa'$}
\DisplayProof

& \quad \quad &

\AxiomC{$\Gamma_j\ \vdash\ M\,:\,A_j\, ::\, \kappa \quad (\forall j\in J)$}
\RightLabel{\quad Promotion} 
\UnaryInfC{$\sum_ {j \in J} \Gamma_j\ \vdash\ M\,:\,[A_j\,\vert\, j \in J]\, ::\, \kappa $}
\DisplayProof
\\
\end{tabular}


$$
\AxiomC{$\Gamma_1\ \vdash\ N_1\,:\, A_1 \,::\, \kappa_1 \quad\ \ \cdots \ \ \quad \Gamma_n\ \vdash\ N_n\, :\, A_n\, ::\, \kappa_n $}
\AxiomC{$\quad \quad \ \ \quad q \in Q$}
\LeftLabel{Left negation \quad} 
\BinaryInfC{$\left(\sum_{i=1}^n\ \Gamma_i \right) ,\, f\,:\, \neg_q\,\left(A_1,\,\cdots,\,A_n\right)\,::\, 
\lnot\, \left(\kappa_1, \dots, \kappa_n\right)\, \vdash\ f\ N_1 \cdots N_n : q :: \bot$}
\DisplayProof
$$


$$
\AxiomC{$\Gamma,\, x_1\,:\,A_1 \,::\, \kappa_1,\, \ldots ,\, x_n\, :\, A_n :: \kappa_n\ \vdash\ M \, :\, q\, ::\, \bot $ }
\AxiomC{$\quad \quad q \in Q$}
\LeftLabel{Right negation \quad} 
\BinaryInfC{$\Gamma\ \vdash\ \lambda x_1 \cdots \lambda x_n .\, M\, :\,\, \neg_q \left(A_1,\,\dots,\,A_n\right)\, ::\, \lnot \, \left( \kappa_1 , \dots , \kappa_n \right)$}
\DisplayProof
$$
\end{center}

\caption{Extension of tensorial logic with intersection types and color modalities (main rules)}
\label{fig/tensorial-calculus}
\end{figure}

\end{document}